\documentclass[11pt]{article}
\usepackage[margin=2.5cm]{geometry}
\usepackage[T1]{fontenc}
\usepackage{amsmath,amssymb}
\usepackage{graphicx}
\usepackage{booktabs,microtype}
\usepackage{placeins,array,longtable}

\graphicspath{{figures/}}
\newtheorem{theorem}{Theorem}
\newcommand{\dd}{\mathrm d}\newcommand{\ii}{\mathrm i}

\newcommand{\gam}{\gamma}\newcommand{\sinc}{\operatorname{sinc}}

\newcommand{\atanh}{\operatorname{artanh}}
\usepackage[colorlinks=true,linkcolor=blue,citecolor=blue,urlcolor=blue]{hyperref}

\begin{document}
\begin{center}
{\Large\bfseries Thermal aliases in quantum noise: sampling clocks and pulse widths}\\[2mm]
Zi-Xuan Wang$^{1,\dagger}$, Xiu-Hao Deng$^{2,3,\dagger,*}$\\[1mm]
{\small $^{1}$Department of Physics, Southern University of Science and Technology, Shenzhen, China}\\
{\small $^{2}$International Quantum Academy, Shenzhen 518048, China}\\
{\small $^{3}$Shenzhen Branch, Hefei National Laboratory, Shenzhen 518048, China}\\[1mm]
{\small $^{\dagger}$These authors contributed equally to this work.\quad
$^{*}$Corresponding author: dengxiuhao@iqasz.cn}
\end{center}

\begin{abstract}
Can collective quantum-noise measurements distinguish different mode temperatures from one common temperature when the coupling spectrum is unknown? For stationary Gaussian oscillator environments, we derive a sharp bandwidth threshold for an exact thermal alias on the integer-time sampling lattice. The threshold depends on temperature order through the quantum response. For frequencies within the first alias period, fixing sufficient integer-time correlations makes one additional fractional-time fluctuation observation enough to exclude every common-temperature fit. A calibrated second pulse width provides another route through the same collective port: both widths must probe one spectrum and retain their measured variance scales. For a specified two-mode example, continuous-frequency certificates cover all common temperatures and uncertainty in a shared pulse ramp. A binary-readout protocol combines randomly interleaved matched references with collective-variance and coherence checks to handle Gaussian oscillator loss and predictable readout drift. The selected fractional-clock and two-width designs require approximately $3.8\times10^7$ and $1.6\times10^8$ outcomes for at most 5\% type-I error and at least 95\% power on the stated alternative class. These theoretical and numerical acquisition guarantees assume independently validated preparation, spectral support and control specifications; validation costs are additional. The result identifies which new observation makes thermal-model testing informative when mode-resolved access is unavailable.
\end{abstract}

\section{Introduction}

A thermal noise model is useful only if its temperature and spectrum describe the observables used by a sensor or simulator. Quantum response contains information that a classical covariance lacks: frequency-resolved fluctuation and response spectra can determine an effective temperature, and a frequency-dependent value diagnoses departure from equilibrium\cite{PazSilva2017,Wang2021}. The difficulty considered here arises when only a few collective time-domain observations are available and the positive coupling spectrum is unresolved. Refitting that spectrum can compensate for changing the temperature. Excluding equilibrium therefore requires testing every common temperature and every allowed positive spectrum. Spectral signatures of engineered nonthermal baths provide related diagnostics under different bath models\cite{Roman2020}.

We ask a concrete experimental-design question: when should a measurement campaign change its sampling times or pulse widths instead of repeating the original settings? Our starting point is a sharp thermal alias boundary. At a fixed pulse filter, an environment with different mode temperatures can be indistinguishable on the entire integer lattice from a physical environment at one common temperature. Within a single alias period, one additional fractional-time cosine removes every such explanation for a finite-mode family. A second pulse width through the same port can also resolve the ambiguity. We compare both routes after pulse averaging, oscillator loss and binary readout statistics.

The mathematical and experimental ingredients have established precedents. Positive spectral moment sets and their extremal measures underlie rigorous spectral uncertainty\cite{Karlsson2013,Yang2018}. Control frames, Bayesian inference and variational noise spectroscopy address useful finite-data characterization\cite{Frame2021,Ferrie2018,Shitara2025}. In particular, the frame example fits two oscillator temperatures and compares coarse and fine time resolution inside a known two-mode model\cite{Frame2021}. Energy--entropy inequalities provide certified bounds for Gibbs-state observables and Hamiltonian learning\cite{Fawzi2024,Artymowicz2024}. Nonuniform sampling in NMR and band-limited or digital-filter quantum noise spectroscopy provide established ways to choose sampling schedules and filters\cite{Maciejewski2009,Norris2018,Digital2024}.

The distinction here is that the null may refit an unknown mode count, continuous frequencies and arbitrary positive variance weights while sharing one unknown temperature. We obtain a temperature-order-dependent \emph{if-and-only-if} bandwidth threshold, rather than suppressing replicas for a fixed spectrum, and then certify a separating observation uniformly over the continuous frequency and temperature domains. The finite-sample layer turns this identifiability result into a conditional binary-readout decision. These are the analytic theorem, numerical two-mode example and acquisition benchmark, respectively; their scopes are separated in table~\ref{tab:claimmap}.

\begin{table}[tbp]
\centering\small
\begin{tabular}{p{.18\linewidth}p{.38\linewidth}p{.35\linewidth}}
\toprule
Level & Established result & Fixed inputs and boundary\\\midrule
Analytic & Exact alias cutoff and one-observation separators for arbitrary positive variance-spectrum refitting & Finite target modes in the first alias period; no two-mode example-specific error model\\[2pt]
Certificate & Positive distance over all common temperatures and continuous frequency/ramp support & The stated two-mode example, band, tail and pulse family\\[2pt]
Acquisition & Conditional size/power guarantee and sufficient binary-outcome budgets & Supplied preparation, loss, control, matching and visibility domain; no claim of hardware demonstration or optimal sample complexity\\
\bottomrule
\end{tabular}
\caption{Claim map. Each later quantitative statement belongs to one of these three levels.}
\label{tab:claimmap}
\end{table}

\section{Noise model and sensor readout}
\label{sec:model}

\begin{figure}[tbp]
\centering\includegraphics[width=.85\linewidth]{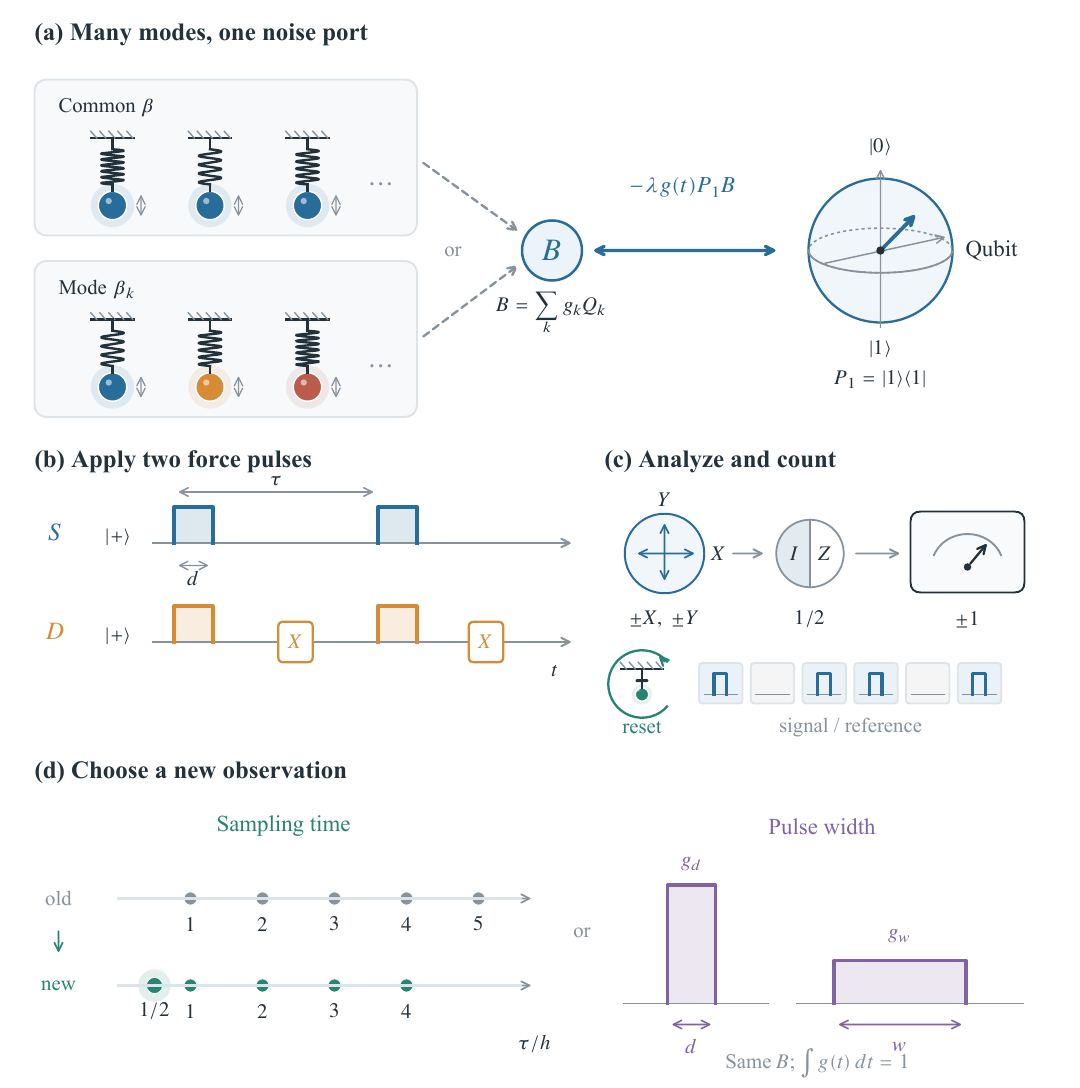}
\caption{Physical model and measurement choices. (a) Independent oscillator modes may share one temperature or have mode-dependent temperatures; the common-temperature model may refit their frequencies and positive variance weights. A qubit probes only the collective coordinate $B$. (b) The paired $S,D$ magnitudes measure fluctuation; the swapped $D$ phase retains response after equal self phases cancel. (c) Randomly interleaved signal/reference slots use four analysis phases, a final fair $I/Z$ choice and binary readout; each slot starts from a fresh prescribed bath state. (d) For the illustrated measurement schedules, the alias is removed either by replacing one integer lag with $0.5h$ or by adding a calibrated wider pulse through the same port. Table~\ref{tab:protocol} states the required premises.}
\label{fig:scheme}
\end{figure}

We use $\hbar=k_B=1$ and a base time interval $h$. The frequency $\omega$ and inverse temperature $\beta$ are dimensionless, with physical Boltzmann exponent $\beta\omega$. If the oscillator Hamiltonian is an effective rotating-frame Hamiltonian, the tested Gibbs temperature is defined with respect to that Hamiltonian and is not obtained by inserting a laboratory carrier frequency. The independent modes in figure~\ref{fig:scheme}(a) have
\begin{equation}
 H_E=\sum_k\frac{\omega_k}{2}(Q_k^2+P_k^2),\quad [Q_k,P_l]=\ii\delta_{kl},\quad B=\sum_k g_kQ_k.
 \label{eq:ham}
\end{equation}
The initial state is a product of centered thermal mode states, possibly at different $\beta_k$. This produces stationary Gaussian correlations even when the product is not Gibbs at one temperature. Let $v=\langle B^2\rangle$, $p_k=g_k^2\coth(\beta_k\omega_k/2)/(2v)$, and $\gam_\beta(\omega)=\tanh(\beta\omega/2)$. The normalized ordered correlation is
\begin{equation}
 k(\tau)=\sum_kp_k\left[\cos(\omega_k\tau)-\ii\gam_{\beta_k}(\omega_k)\sin(\omega_k\tau)\right].
 \label{eq:kernel}
\end{equation}
We write $x=\Re k$ and $y=-\Im k$. The latter is proportional to the commutator, rather than an additional classical covariance. In the convention $\mu=\langle[B(t),B(s)]\rangle/(2\ii)$, the susceptibility to a perturbation $-F B$ is $\chi=-2\Theta\mu$.

The thermal null is the family we try to exclude. Each member has one inverse temperature $\beta>0$ and one positive probability measure $q$ on a supplied band $[a,b]$. The same temperature and measure must explain every observation. The number of modes, frequencies and coupling weights may all be refitted.

We call $v\,\dd q$ the positive variance spectrum. It weights each mode by its contribution to $\langle B^2\rangle$, including thermal occupation. At fixed $v$, every positive atom is physically realized by $g_k^2=2vp_k\gam_\beta(\omega_k)$. Thus a temperature change can be compensated by changing the couplings while retaining the thermal relation between fluctuation and response.

The band is supplied independently. In the finite-error example, a supplied fraction $\ell$ of the initial bare variance may lie outside it. These support bounds, as well as any known mode frequencies, force phase or mode-resolved control, are shared inputs to all comparators.

The sensor applies a state-selective force: only its $|1\rangle$ arm drives the collective coordinate $B$. With $P_1=|1\rangle\langle1|$,
\begin{equation}
 H_c(t)=-\lambda g_d(t)|1\rangle\langle1|B,\qquad \lambda\ge0,
 \label{eq:force}
\end{equation}
where $g_d=1/d$ on a box of width $d$, so each pulse has unit area and $\lambda^2v_g$ is dimensionless. Between two boxes separated by $\tau>d$, an $X$ rotation exchanges the sensor populations and selects which arm experiences the second force. Restoring the sensor labels at the end gives two histories: same arm, $(h_1,h_0)=(e+l,0)$, and different arms, $(e,l)$, where $e=g_d(t)$ and $l=g_d(t-\tau)$. We denote these protocols by $S,D$. The identity $X|1\rangle\langle1|X=|0\rangle\langle0|$ retains the unconditional force required by the response phase. Internal-state-dependent impurity coupling provides a theoretical route to this interface\cite{Klein2007}. Applying it to a harmonic noise module also requires reproducible mode occupations, the same collective coordinate at both pulse widths, and validated switching and rotation errors. These requirements are summarized in Table~\ref{tab:protocol}.

Put $G_d(\omega)=\sinc(\omega d/2)$, $v_g=\int G_d^2v\dd q$ and $\dd q_g=G_d^2v\dd q/v_g$. Write $x_g,y_g$ for equation~\eqref{eq:kernel} with $q_g$, and let $S_d=\sum_kv_k\gam_k(\omega_kd-\sin\omega_kd)/(\omega_kd)^2$. The exact coherent sensor signals are
\begin{align}
 W_S(\tau)&=e^{-\lambda^2v_g(1+x_g)}e^{\ii\lambda^2(2S_d+v_gy_g)},\nonumber\\
 W_D(\tau)&=e^{-\lambda^2v_g(1-x_g)}e^{-\ii\lambda^2v_gy_g}.
 \label{eq:coherence}
\end{align}
The magnitudes of $S$ and $D$ determine the filtered fluctuations; the $D$ phase measures the ordered cross response after the equal self phases cancel. We use only this phase. A stable phase per occupied force box also cancels in $D$, whereas an early--late mismatch enters the error budget. For this readout, the force has a nonzero sum across the two arms. A symmetric opposite-arm $\sigma_z$ force has zero sum and loses the corresponding response phase (Supplement S1).

Each signal has a force-off reference with the same sensor controls and duration. Four analysis phases recover the two coherence quadratures. An independent fair $I/Z$ choice, applied after the analysis rotation and before binary detection, removes transverse detector sensitivity; opposite phases cancel the detector offset. The matched reference then removes the common gain and sensor phase. Related model-free readout mitigation uses randomization and calibration to remove multiplicative detector bias\cite{Berg2022}. Our procedure requires trusted rotations and a detector map shared by the two possible force choices within each slot. Between slots the map may drift. Supplement S4 gives the detector algebra; Section~\ref{sec:counts} describes acquisition during drift.

When $G_d$ has no zeros on the band, every positive filtered spectrum can be produced by a positive unfiltered spectrum. The band-limited thermal geometry therefore survives finite pulse width. Its variance scale, self phase and out-of-band contribution still require the separate accounting below.

\FloatBarrier
\section{A sharp thermal alias boundary}
\label{sec:alias}

Take $M\ge2$ target frequencies $0<\omega_1<\cdots<\omega_M<\pi$, positive weights, and finite $\beta_i>0$ that are not all equal. Let $\beta_* =\max_i\beta_i$, the inverse temperature of the coldest mode, and $J=\{i:\beta_i<\beta_*\}$. The following theorem identifies an exact threshold even though the null has no fixed mode count.

\begin{theorem}[Thermal cutoff and one fractional time]
\label{thm:clock}
Let $0<a\le\omega_1$, $\omega_M\le b<2\pi$. A common-temperature positive variance spectrum on $[a,b]$ matches every integer-time correlation of the target if and only if
\begin{equation}
 b\ge b_c=\max_{i\in J}(2\pi-\omega_i)
             =2\pi-\min_{i\in J}\omega_i.
 \label{eq:cutoff}
\end{equation}
Matching only $x(1),\ldots,x(2M)$ and $y(1),\ldots,y(M)$ is already equivalent to this condition. Adding $x(r)$ at any $0<r<1$ excludes every such common-temperature spectrum for every admissible $b<2\pi$. Below $b_c$, or after this addition, the corresponding finite vector has strictly positive distance from the all-temperature closure.
\end{theorem}

The alias has a simple origin. At every integer lag, a frequency $\omega_i$ and its reflection $2\pi-\omega_i$ have the same cosine but opposite sine. Moving positive spectral weight between them leaves fluctuations unchanged and adjusts the response. A sufficiently cold common-temperature environment can therefore mimic a hotter mode by assigning weight to its reflected partner. Positivity determines which partners are necessary and hence how wide the allowed band must be.

To make this argument precise, set $c_i=\cos\omega_i$. The nonnegative cosine polynomial
\begin{equation}
 Q(\omega)=\prod_{i=1}^M(\cos\omega-c_i)^2
 \label{eq:annihilator}
\end{equation}
has degree $2M$ and zero target expectation. Matching those real moments forces every null atom onto $\omega_i$ or its reflection $\Omega_i=2\pi-\omega_i$, when allowed by the band. A polynomial-basis Vandermonde matrix fixes each pair's total mass to $p_i$. The first $M$ sine moments then fix its net response coefficient. If $s_i$ is the reflected mass, it obeys
\begin{equation}
 p_i\gam_{\beta_i}(\omega_i)
 =(p_i-s_i)\gam_\beta(\omega_i)-s_i\gam_\beta(\Omega_i)
 \le p_i\gam_\beta(\omega_i).
 \label{eq:responsepair}
\end{equation}
Thus $\beta\ge\beta_*$, and every hotter mode needs a reflected partner. This proves the necessity of equation~\eqref{eq:cutoff}. At $\beta=\beta_*$, split its mass using
\begin{equation}
 \theta_i=\frac{\gam_{\beta_i}(\omega_i)+\gam_{\beta_*}(\Omega_i)}
 {\gam_{\beta_*}(\omega_i)+\gam_{\beta_*}(\Omega_i)},\qquad
 (q_i,q_{\Omega_i})=(p_i\theta_i,p_i(1-\theta_i)).
 \label{eq:split}
\end{equation}
The coldest modes remain unchanged. This constructs at most $2M-1$ physical modes at one temperature, matching all integer lags.

The new real observation has a definite sign:
\begin{equation}
 x_{\rm target}(r)-x_{\rm null}(r)
 =2\sin(\pi r)\sum_i s_i\sin[r(\pi-\omega_i)].
 \label{eq:fractional}
\end{equation}
All factors multiplying the reflected masses are positive. Equality forces all $s_i=0$, after which equation~\eqref{eq:responsepair} demands incompatible temperatures. The $\beta\to0,\infty$ limits also fail. They are included as closure endpoints when proving a positive distance, not as additional finite-temperature nulls. Compactness yields the positive gap; its finite-error size can shrink as the new time approaches an integer, temperatures merge or frequencies coalesce. Supplement S2 gives the proof and degeneracies. This finite set of correlations supplies a sufficient test.

The cutoff depends on temperature order. For frequencies $(0.7,1.6,2.2)$, assigning inverse temperatures $(1.7,0.3,1.7)$ gives $b_c=2\pi-1.6$, whereas $(0.3,1.7,1.7)$ gives $b_c=2\pi-0.7$. At $b=5$, the former has an integer-time thermal alias and the latter does not. This is a thermally constrained distinction beyond the location of a classical sampling replica.

Our quantitative two-mode example is
\begin{equation}
 \omega=(0.7,1.6),\qquad p=(1/2,1/2),\qquad \beta=(1.7,0.3),
 \label{eq:fixture}
\end{equation}
with initial bare variance $v=1$. Its cutoff is $4.683185\ldots$.

\noindent\begin{minipage}{\linewidth}
In a nonfolding band one complex lag always admits a one-mode thermal explanation, but two can suffice to exclude every common temperature. On $[0.05,2.5]$, independently verified point-observation distance brackets are
\begin{center}
\begin{tabular}{ccc}\toprule
Complex lags & Certified lower & Physical upper\\\midrule
$1,2$ & $0.0216294$ & $0.0274257$\\
$1,2,3$ & $0.0985292$ & $0.1057936$\\
$1,2,3,4$ & $0.1324644$ & $0.1467426$\\\bottomrule
\end{tabular}
\end{center}
\end{minipage}

For the wider band $[0.05,5]$ and width $d=0.2$, five integer lags retain an exact filtered alias. The schedules $(0.5,1,2,3,4)$ and $(0.5,1,1.5,2)$ have certified lower distances $0.0649854$ and $0.00209880$, respectively. A separate physical four-atom construction at common $\beta=13/20$ bounds the latter distance above by $0.04050609$, strictly below the former's lower bound. This orders their true normalized distances for these different maps, without implying a resource advantage. Supplement S3 gives the exact positive atoms, temperature leaves and constructions behind both broad- and narrow-band brackets.

\begin{figure}[tbp]
\centering\includegraphics[width=0.8\linewidth]{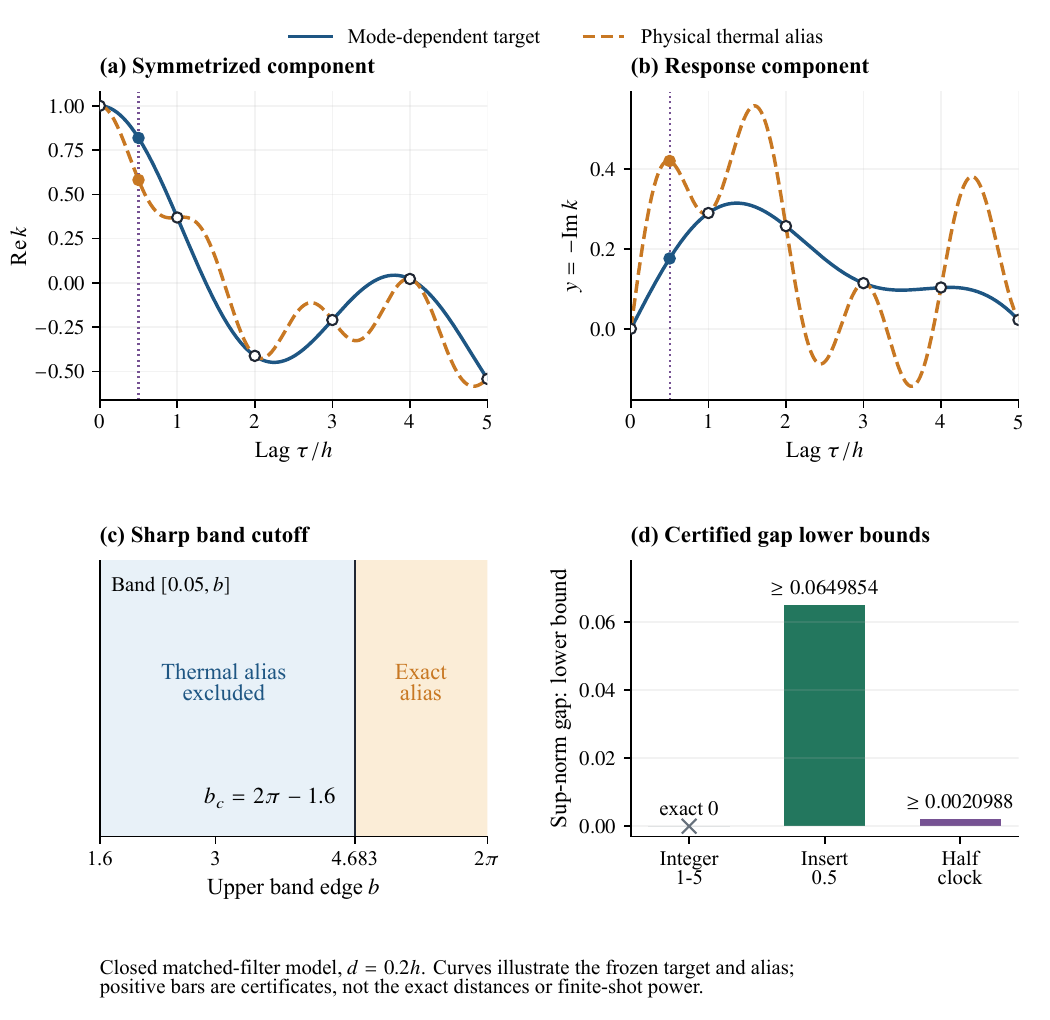}
\caption{Thermal aliasing and the information supplied by a fractional time. (a,b) The width-$0.2$ filtered correlations of the two-mode example and its three-mode common-$\beta=1.7$ alias agree at every integer lag in fluctuation $x_g$ and response $y_g$. A fractional-time observation separates them. (c) The cutoff $b_c=2\pi-1.6$ lies inside the illustrated band ending at $5$; the boundary admits an alias. (d) Certified lower distances for the indicated observation maps on the continuous band $[0.05,5]$. A physical construction gives the four-lag map an upper distance $0.04050609$, below the five-lag map's lower bound $0.0649854$. The integer map has zero distance by construction. These distances describe geometric separation; acquisition costs also depend on readout and systematic errors.}
\label{fig:alias}
\end{figure}

\FloatBarrier
\section{Excluding all common-temperature spectra}
\label{sec:certificate}

For selected times, collect the fluctuation coordinates $x_g$ followed by the response coordinates $y_g$ into $Y$. A witness $z$ selects a linear combination of these correlations. We bound the largest value that a thermal spectrum can give to that combination. If the measured value exceeds the bound, the thermal explanation is excluded. The distance $\|Y-Y'\|_\infty$ is the largest coordinate difference. At fixed $\beta$, the allowed vector is in the convex hull of
\begin{equation}
 b_\beta(\omega)=\big((\cos\omega\tau_j)_j,
              (\gam_\beta(\omega)\sin\omega\tau_j)_j\big).
 \label{eq:atom}
\end{equation}
The union over temperatures is generally nonconvex. The null uses the same temperature at every frequency; mixing temperatures enlarges that hypothesis class. For a real direction $z$ with $\|z\|_1\le1$, let $H_\beta(z)=\sup_{\omega\in[a,b]}z\cdot b_\beta(\omega)$. A useful identity is
\begin{equation}
 \sup_{\beta\in[\beta_L,\beta_R],\,\omega\in[a,b]}z\cdot b_\beta(\omega)
 =\max\{H_{\beta_L}(z),H_{\beta_R}(z)\}.
 \label{eq:endpoints}
\end{equation}
At fixed $\omega$, the expression is affine in the monotone function $\gam_\beta(\omega)$, so its maximum lies at a temperature endpoint even when its sine coefficient changes sign. The endpoints $0$ and $\infty$ are included exactly. We cover $u=\beta/(1+\beta)\in[0,1]$ by finitely many intervals, each with its own witness. These intervals are called leaves in the certificate tables. Excluding every leaf excludes every common temperature while preserving the shared-temperature constraint.

Frequency remains continuous. For a grid of maximal spacing $\Delta\omega$, an outward upper bound is the largest interval-evaluated node value plus $L\Delta\omega/2$, with
\begin{equation}
 L=\sum_j|z_{R,j}|\tau_j+|z_{I,j}|\left(\tau_j+\min\{\beta/2,1/a\}\right).
 \label{eq:lip}
\end{equation}
At the closure endpoints $\beta=0,\infty$ the derivative term is zero. Interval arithmetic rounds lower endpoints down and upper endpoints up. Together with exact rational directions and norm checks, it makes a positive minimum margin a lower bound on the all-temperature sup-norm distance. Grid-based positive spectral fits separately give physical upper bounds. The supplementary data contain the exact witnesses, endpoint-node records and a fixed verification entry point (Supplement S3).

For bare tail fraction $\ell$, put $m=\min_{[a,b]}G_d^2>0$. The filtered tail is at most $\ell_g=\ell/[m(1-\ell)+\ell]$. Since every coordinate in equation~\eqref{eq:atom} is bounded by one, a common valid support is
\begin{equation}
 H=(1-\ell_g)\min\{H_{\rm band},\|z\|_1\}
                         +\ell_g\|z\|_1.
 \label{eq:tail}
\end{equation}
The same out-of-band mass enters every observation. Self-phase observables require a different tail treatment because filter zeros can hide variance while leaving a self phase.

\paragraph{Computational workflow.}
Floating-point optimization proposes witness directions and count allocations.
These are fixed before count generation and checked using outward interval
arithmetic. The supplementary data provide the exact inputs, witness records,
node bounds and verification code (S3 and S9).

\FloatBarrier
\section{A finite-sample thermal test}
\label{sec:counts}

The test concerns the mode temperatures immediately after each fresh bath preparation. The initial state, spectrum and pulse settings are fixed across the campaign. Loss rates and reservoir occupations may change between preparations in response to earlier records, but are chosen before the next random force label. They remain constant within each coherent slot and are shared by its two possible force choices. Table~\ref{tab:protocol} collects the physical inputs and Table~\ref{tab:decisionflow} gives the sequence of measured checks and the final decision.

\begin{table}[!htbp]
\centering\scriptsize
\begin{tabular}{p{.19\linewidth}p{.22\linewidth}p{.50\linewidth}}
\toprule
Requirement & Provenance & Role in the final decision\\\midrule
Harmonic module, band and tail & Independently specified or calibrated & One unresolved collective port; fresh prescribed preparation; $[.05,5]$ and bare-tail fraction $\ell\le10^{-3}$. These are not inferred from the thermal-test data.\\[2pt]
New observation & Protocol controlled & Use width $.2$ at lags $(.5,1,2,3,4)$, or widths $.2,.8$ at integer lags $1,2,3,4$. Timing is validated for the first route; common port, area and ramp are validated jointly for the second.\\[2pt]
Signal--reference matching & Independently validated premise & A dummy-waveform, detuned or zero-coupling control must bound label-dependent complex-mean mismatch within the residual envelope and state its validity interval. Failure or expiry returns inconclusive.\\[2pt]
Binary acquisition & Protocol controlled & Predetermine every stream count and coefficient; use $\lambda^2=1.1$, force probability $p_A=3/4$, four analysis phases, a fair final $I/Z$ randomization and fresh preparations for signal and reference slots.\\[2pt]
Variance, loss and visibility & Measured gate plus calibrated caps & An 80,000-slot single-box collective-variance stage bounds $v$ by $2$; loss, control and nonzero-reference checks gate use of the certificate.\\[2pt]
Decision and cost boundary & Theorem plus confidence bounds & Reject only when all gates pass and both temperature intervals separate. Conditional type-I error is at most $.05$; specification, timing and width validation, with their failure allocation, are additional costs. No complete hardware module is demonstrated here.\\
\bottomrule
\end{tabular}
\caption{Feasibility and provenance map for the two selected acquisition designs. The null has one unknown common temperature and an arbitrary positive variance spectrum.}
\label{tab:protocol}
\end{table}

\begin{table}[!htbp]
\centering\small
\begin{tabular}{p{.20\linewidth}p{.70\linewidth}}
\toprule
Step & Operation and outcome\\\midrule
1. Validate inputs & Confirm the supplied support, preparation and matching specifications and their validity interval.\\[2pt]
2. Acquire counts & Use the predetermined stream lengths and randomized force/reference labels. Estimate each complex mean and its error disk.\\[2pt]
3. Check observables & Apply the collective-variance bound, positive-modulus checks and the $D$ phase-branch check. A failed check returns inconclusive.\\[2pt]
4. Compare scores & For each temperature interval, compare $\widehat F_z$ with the tail and measurement allowance $B_z+E_z$.\\[2pt]
5. Decide & Reject the common-temperature model only if every interval separates. Otherwise return inconclusive.\\
\bottomrule
\end{tabular}
\caption{Fixed-length thermal-model test. The confidence bounds control the false-rejection probability conditional on the validated physical specifications. Rejection tests those premises jointly with the common-temperature hypothesis; an inconclusive result leaves that hypothesis unresolved.}
\label{tab:decisionflow}
\end{table}

\FloatBarrier

\subsection{Preparation, loss and a collective-variance check}

Time zero is completion of bath preparation. The first force follows the initial Ramsey rotation after $t_{\rm pre}\le.01h$. During that coherent slot, each oscillator undergoes linear Gaussian loss,
\begin{equation}
 \dot\rho=-\ii[H_E+H_c,\rho]
 +\sum_k\kappa_k[(n_{b,k}+1)\mathcal D[a_k]+n_{b,k}\mathcal D[a_k^\dagger]]\rho.
 \label{eq:loss}
\end{equation}
Here we suppress the slot index. The caps are $\kappa_kh\le\bar\kappa$ and $0\le n_{b,k}\le1/2$. The covariance need not be stationary: the reservoir occupations can differ from the initial mode occupations. The commutator still depends only on the time difference, so the two $D$ self phases cancel.

Before using a loss allowance, we bound the initial collective variance $v=\langle B^2\rangle$ with a unit-strength single-box measurement. This measures the driven collective coordinate, rather than the total bath energy. Let $m_{\rm filt}$ be a known lower bound on the squared pulse filter over the supplied band and waveform family. For the finite-ramp pulses used here, the box bound $m_{\rm filt}=\min_{\omega\in[a,b]}G_d^2(\omega)\simeq.9194$ is valid. The per-slot signal factor satisfies $|W_{E,i}|\le e^{-K_{\min}v/2}$, where $K_{\min}=m_{\rm filt}(1-\ell)-\bar\kappa(d+2t_{\rm pre})/2>0$.

The binary-detector reference obeys $|B_i|\le1$. Hence the ideal collective-variance conditional mean satisfies $|A_E^*|\le e^{-K_{\min}v/2}$ even when loss and reference phase vary between slots. If $\widehat A_E$ estimates this mean with error radius $e_E$, define
\begin{equation}
 a_L=|\widehat A_E|-e_E,\qquad
 v_{U,\rm abs}=-2\log(a_L)/K_{\min}.
 \label{eq:abs-main}
\end{equation}
Continue only if $0<a_L\le1$ and $v_{U,\rm abs}\le2$. A passing check bounds the variance; it does not itself reject the thermal model. Using the absolute magnitude avoids taking a ratio of averages with different per-slot loss factors. Supplement S6 derives this bound and the radius used for the 80,000-slot stage.

With $v\le V_{\max}=2$, the complex-coherence deviation from the closed signals in equation~\eqref{eq:coherence} is bounded by
\begin{equation}
 \epsilon_{{\rm loss},j}^{\rm pre}(V_{\max})=\lambda^2V_{\max}\frac{\bar\kappa}{2}
 \sqrt{4(\tau_j+d+2t_{\rm pre})^2+(\tau_j+d/3)^2}.
 \label{eq:lossbound}
\end{equation}
The preparation interval enters the covariance term; the stationary response term is unchanged. Supplements S4 and S6 derive the fixed-loss kernel and this preparation-origin bound. The reservoir-occupation cap belongs to the finite-error example. A larger known cap increases the fluctuation allowance as specified in S4. The acquisition design assumes Gaussian linear loss.

\subsection{From coherences to a thermal score}

Write the closed comparison signals and references as $A_{j,q}=B_{j,q}W_q$, $q=S,D$, with matching complex attenuation in each pair. The two protocols may have different reference phases. For $C_q=A_q/B_q$ define
\begin{align}
 u_j&=\log|C_{D,j}|-\log|C_{S,j}|,\nonumber\\
 t_j&=-\log|C_{D,j}|-\log|C_{S,j}|,\qquad
 p_j=-2\arg C_{D,j}.
 \label{eq:rawstats}
\end{align}
On the unwrapped phase branch, equation~\eqref{eq:coherence} gives $(u_j,t_j,p_j)=2\lambda^2v_g(x_g(\tau_j),1,y_g(\tau_j))$. Thus $u_j$ measures fluctuation, $t_j$ sets its variance scale, and $p_j$ measures response. For each fixed witness and nonnegative normalization weights $a_j$ summing to one, the score is
\begin{equation}
 F=\sum_j(z_{R,j}u_j+z_{I,j}p_j-Ha_jt_j)
   =2\lambda^2v_g(z\cdot Y-H).
 \label{eq:score}
\end{equation}
A positive score exceeds the thermal support bound. In measured data, the variance, modulus and phase checks ensure that the loss allowance, logarithms and response branch are valid. The confidence bounds below quantify the remaining uncertainty.

\subsection{Acquiring a common reference during drift}

A reference measured before a long campaign need not describe later signals.
Drift-aware quantum characterization and SPAM-robust spectroscopy already
address this general difficulty\cite{Proctor2020,Khan2024}. Here a matched
randomized schedule provides the common reference required for the thermal decision.
For each protocol and lag, fix $n$ equal-duration slots in advance. Every slot
requires a complete bath preparation, including force-off references. Draw a fresh
force-on label $I_i$ with probability $p_A$ before coherent sensor preparation.
After evolution, draw a uniform analysis phase
$\theta_i\in\{0,\pi/2,\pi,3\pi/2\}$, then apply the analysis rotation, an independent
fair $I/Z$ coin and binary readout $s_i\in\{-1,1\}$.

Before the fresh choices, let the conditional signal and reference coherences
be $B_iW+\delta_{A,i}$ and $B_i+\delta_{B,i}$, with $|B_i|\le1$ and bounded
residuals. The common factor and detector offset may depend on previous data.
The two possible force choices must share this background in each slot. The
closed environment and its pulse shape are fixed across the campaign. The
four-phase conditional mean then gives the estimators
\begin{equation}
 \widehat A=\frac1n\sum_{i=1}^n\frac{2I_is_i e^{\ii\theta_i}}{p_A},\qquad
 \widehat B=\frac1n\sum_{i=1}^n
             \frac{2(1-I_i)s_i e^{\ii\theta_i}}{1-p_A}.
 \label{eq:randommeans}
\end{equation}
Their ideal predictable averages are $A^*=\overline BW$ and
$B^*=\overline B$, where $\overline B=n^{-1}\sum_iB_i$ can itself be random.
The conditional offset cancels at each slot. Equation~\eqref{eq:randommeans}
uses the prespecified total $n$, not the realized number of force-on labels.
It estimates the same averaged reference for signal and force-off data during drift.

Random interleaving shares the averaged drift between signal and reference.
It still requires their two possible force choices to be matched within a slot.
Before acquisition, an independent placebo sequence---for
example a dummy waveform, a far-detuned force or a zero-coupling control---must
bound label-dependent phase, heating and detector changes without using the
unknown spectrum or temperature. Its complex-mean bound is charged to the
residual envelope; its counts, duration, confidence allocation and validity
interval belong to $T_{\rm spec}$ and $\alpha_{\rm spec}$. If this check fails
or expires, the reported outcome is inconclusive rather than nonthermal.

Each real or imaginary inverse-probability variable has conditional second
moment $2/p_A$ and centered increment bounded by $4/p_A$. A conservative
fixed-horizon Bernstein--martingale bound\cite{Freedman1975} therefore gives
the complex radius
\begin{equation}
 e(n,p,L)=\sqrt{\frac{8L}{pn}}+\frac{8\sqrt2L}{3pn}.
 \label{eq:martingale}
\end{equation}
For $m$ paired lags, use $L_A=\log(8m/\alpha_A)$ and
$L_B=\log(8m/\alpha_B)$ for the $2m$ signal and $2m$ reference
channels. More generally, for $N$ retained complex streams use
$L_A=\log(4N/\alpha_A)$ and $L_B=\log(4N/\alpha_B)$,
with the prescribed horizon of each stream. Signal radii include the
statistical term, residual cap and loss allowance; reference radii include
the statistical term and residual cap. The separate 80,000-slot
collective-variance stage uses $p_E=1/2$ and $L_E=\log(8/\alpha_E)$.
Its measured bound determines whether the downstream loss allowance
can be used. Its bound and the final allocations are given in S6 and S9.

Logarithms convert coherence magnitudes to fluctuations and the $D$ phase to response. To control this transformation near zero, for every observed complex mean $\widehat X$ require
$|\widehat X|>e_X$ and put $d_X=-\log(1-e_X/|\widehat X|)$. Define
\begin{align}
 c_{D,j}&=\sqrt{(z_{R,j}+Ha_j)^2+4z_{I,j}^2},&
 c_{S,j}&=|-z_{R,j}+Ha_j|,\nonumber\\
 E_z&=\sum_j\big[c_{D,j}(d_{AD,j}+d_{BD,j})
                    +c_{S,j}(d_{AS,j}+d_{BS,j})\big].
 \label{eq:diskallowance}
\end{align}
For the final comparison, $H$ is the band support. Before acquisition we expand $F$ into coefficients multiplying each measured $S$ log-magnitude and $D$ complex logarithm. A stream is omitted only if its coefficient is exactly zero in both temperature intervals and no retained gate requires it. The expanded expression is then the measured score; individual $u_j$ and $t_j$ need not be reconstructed from an omitted stream. The common bare-tail allowance is
$B_z=2\lambda^2V_{\max}\ell(\|z\|_1+|H|)$, with $V_{\max}=2$.
After the absolute collective-variance check, $|\lambda^2R_w|\le2\lambda^2<\pi$.
The observed $D$ phase therefore has the correct branch whenever
\begin{equation}
 d_{AD,j}+d_{BD,j}<\pi-2\lambda^2
 \label{eq:phasegate}
\end{equation}
for every retained $D$ stream. This check uses only measured $D$ means.
Reject only when the collective-variance, modulus and phase checks pass and
$\widehat F_z>B_z+E_z$ on both temperature intervals.
Otherwise the result is
inconclusive. With $(\alpha_A,\alpha_B,\alpha_E)=(.02,.02,.01)$, a true
common-temperature model is rejected with probability at most $.05$, given
the physical specifications. A separately estimated specification failure
probability $\alpha_{\rm spec}$ adds to this bound unless allocated beforehand.
The confidence bounds apply to the conditional averages of the observations; the confidence-disk calculation then bounds the logarithmic score on each realized path. A small or vanishing averaged reference is handled by the gates,
without forced division or postselection.

\subsection{A finite budget with a nonzero power domain}

Keep $[a,b]=[.05,5]$, $d=.2$, bare tail $\ell=10^{-3}$, $V_{\max}=2$,
$\bar\kappa=8\times10^{-5}$ and residual complex-mean cap $\epsilon=10^{-3}$.
For power, require on every admissible path
$|\overline B_j|\ge.9e^{-(\tau_j+d)/8}$ and the analogous single-box reference floor.
The corresponding ideal closed-system preparation obeys
\begin{equation}
 \|Y-Y_0\|_\infty\le10^{-3},\qquad
 |v_g-v_{g,0}|\le10^{-3},\qquad v\le1.01,
 \label{eq:powerclass}
\end{equation}
where $v_{g,0}\simeq0.99493$ and $Y_0$ is the finite-box two-mode
example; the design uses the unrounded center. Individual loss rates and bath occupations range over their stated
caps; residual errors may be arbitrary within their bounds.

The selected separate-stream designs and their budgets are compared in
Table~\ref{tab:widthcost}. This is a power statement for alternatives near
the specified target, with nonzero averaged reference visibility; thermal
nulls remain unrestricted by that target neighborhood. The selected allocation
uses $p_A=3/4$, assigning more slots to the signal while retaining enough
force-off references. Raising $p_A$ further increases the reference confidence
radius through $1-p_A$. Deterministic pairing requires matching between
adjacent slots; randomized labels require the two potential force choices to
share a background within each slot.

Saved binary-count diagnostics exercise the final finite-trapezoid readout under
changing matched phase, detector gain and offset. Target datasets reject,
thermal controls remain inconclusive and deliberately vanishing references
fail the observable gates. An independently written implementation reproduces
every recorded decision. These diagnostics test the count interface; the
deterministic bounds, rather than diagnostic frequencies, supply the uniform
size and power guarantees.

\FloatBarrier
\section{Choosing the next measurement}
\label{sec:practice}

\paragraph{Decision order.}
Use a mode-resolved red/blue thermometer when isolated modes and their frequencies are available. If the scientific task concerns one directly accessible downstream observable, measure that observable instead. The collective certificate is appropriate when those routes are unavailable and the question itself is whether \emph{any} common-temperature positive variance spectrum explains an unresolved collective port.

The concrete application is characterization of a prepared harmonic noise module through that port. A band/tail bound, harmonic dynamics, reset procedure and control-error envelope are supplied from its construction or prior validation. The decision is whether the original integer-time acquisition can test a common-temperature explanation, or whether new timing or pulse-filter information is needed. For the two-mode example, repeating the fixed-width integer settings cannot resolve the ambiguity in the declared error class. A fractional time or a jointly constrained second width can. Rejection means that at least the common-temperature explanation or another maintained physical premise fails; inconclusive data do not certify a Gibbs state.

Within the declared setting-dependent residual class, a common-temperature
alias at the fixed-width integer clock matches
the filtered correlations and their scale. An unknown stable phase per
force box can also match the unused self phases. At finite loss the
ideal signals differ slightly. The supplement bounds that difference
and includes the initial Ramsey interval for a finite-ramp example
with $\kappa h=10^{-6}$.

In the full setting-dependent residual-mean envelope of radius $10^{-3}$,
the thermal model can absorb this difference. The two hypotheses then
give the same binary outcome law for each trial, including its random
choices. A test with uniform type-I error at most $.05$ therefore has
power at most $.05$ on the paired target, at any number of repetitions.
This statement covers fresh-reset acquisitions with the original controls,
including adaptive classical processing. It relies on the full declared
residual envelope; it is a minimax statement over that class, and a narrower calibrated error model can add information.

A new fractional time has the positive-power guarantee stated above.
Frequency-resolved fluctuation/response measurements can also distinguish
the temperatures\cite{Wang2021}. Sequential weak-measurement spectroscopy
retains an environment between readouts\cite{Jin2026}, introducing
backaction and bath reuse beyond the fresh-reset comparison considered here.

Mode-resolved sideband thermometry provides a standard baseline\cite{Turchette2000,Vybornyi2023}. For separately addressed isolated thermal modes, matched red/blue transition pairs obey $P_{{\rm red},i}=e^{-\beta_i\omega_i}P_{{\rm blue},i}$ even at finite pulse area, as derived for reciprocal transitions in Supplement S10. With an unknown common detector gain and offset, a measured zero-excitation reference and the same nominal gain lower bound, five binary settings suffice to construct simultaneous temperature intervals. The two-mode example needs about $1.1\times10^4$ preparations under the specified frequency and readout errors (Supplement S10). This much cheaper route requires individual addressing and independently known mode frequencies. Global many-spin excitation and nonthermal single-mode populations require different sideband models\cite{Vybornyi2023,Rasmusson2021}.

Every signal and reference slot requires a fresh sensor and bath preparation and executes the same padded controls. Let $C_{\rm probe}=C_{\rm bath}$ count preparations, $C_{\rm box}$ force boxes, and $C_{\rm swap}$ sensor population flips. Each slot also includes two Ramsey rotations, the force and analysis choices, and the final randomized $Z$ primitive. Exact totals for the final designs are printed in the supplement's final-protocol tables.

Write $t_{\rm probe}$ and $t_{\rm bath}$ for the preparation times,
$t_R$, $t_X$ and $t_Z$ for the Ramsey, population-flip and readout times,
and $t_{\rm settle}$, $t_{\rm switch}$, $t_{\rm force\ label}$ and
$t_{\rm analysis\ label}$ for settling, switching and random-choice times.
A conservative total is
\begin{align}
 T_{\rm total}\le{}&T_{\rm window}+C_{\rm probe}(t_{\rm probe}+t_{\rm bath}
 +2t_R+t_Z+t_{\rm force\ label}+t_{\rm analysis\ label})\nonumber\\
 &+C_{\rm swap}(t_X+t_{\rm settle})+C_{\rm box}t_{\rm switch}+T_{\rm spec}.
 \label{eq:time}
\end{align}
Counting operations already inside a free gap or pulse makes this a conservative
upper bound on time. $T_{\rm spec}$ includes construction, preparation and
prior control-validation costs, with their confidence and validity interval.
All comparators receive the same physical inputs and available metadata.
The reported acquisition is an added acceptance experiment on an already
specified module.

\subsection{Finite pulses and physical time scales}
\label{sec:joint}

A positive trapezoid $f_r=u_{d-r}*u_r$, where $u_L$ is uniform on
$[0,L]$, gives a finite-ramp force of width $d=.2$ and
$r\in[.008,.012]$. Its real filter is
\begin{equation}
 G_r(\omega)=\sinc[\omega(d-r)/2]\sinc(\omega r/2).
 \label{eq:trapezoid}
\end{equation}
On the supplied band $G_r\ge G_d>0$. Outward bounds on the
continuous ramp and common area factor $[.9999,1.0001]$ place the
target family inside the moment and variance domain in
equation~\eqref{eq:powerclass}; the supplement checks the joint-width domain.

The finite-control example uses $t_X=.05h$, $t_R=t_Z=.01h$,
settling time $.02h$, probe preparation $10h$ and bath preparation
$50h$. Its initial Ramsey starts immediately after bath preparation.
With detector contrast at least $.9$, pure Markovian sensor $Z$
dephasing with $T_2\ge10^4h$, and a matched reference phase in
$[\phi_0-.2,\phi_0+.2]$, the averaged visibility satisfies the following bound. Let $T_j$ be the dimensionless coherent-window duration: $T_j=\tau_j+d_j$ for a two-pulse sequence and $T_j=d_j$ for a single pulse, with times expressed in units of $h$. Then
\begin{equation}
 |\overline B_j|\ge .9e^{-(T_j+.18)/10^4}\cos(.2)
                         >.9e^{-T_j/8}.
 \label{eq:cone}
\end{equation}
The $.18h$ overhead includes rotations and analysis-choice latency.
For the shortest window $T=.2$, the sufficient contrast threshold is
about $.8957$; the finite-control error allowance additionally requires
$T_2>7.1\times10^3h$ at the specified unitary-error cap.
The supplement gives both requirements and their sensitivity.

As a conditional accounting example, if $h=1\,\mu\mathrm s$, the selected acquisition times in
Table~\ref{tab:widthcost} become about 40 minutes, 2.76 hours and
10.74 hours before specification, timing and width validation. The same choice
requires 10 ns Ramsey rotations, 50 ns population flips, 8--12 ns
ramps and $T_2\ge10$ ms. A laboratory realization must meet this full set
of requirements at the same $h$. The repeated preparation includes the target's
different mode occupations; extra force-dependent heating must stay
inside the residual allowance. Supplement S5--S6 gives the complete control
specification and includes loss during the initial Ramsey.

\subsection{Keeping the integer clock and changing pulse width}
\label{sec:width}

A tunable force width provides a stronger comparator through the same
collective port. Varying known filters to interrogate one noise spectrum is
established practice\cite{Frame2021,Norris2018,Digital2024}. Here it must test
one common temperature and one common \emph{unfiltered fluctuation-variance
measure}, $\dd\mathcal V=v\dd q$, across both widths. This measure contains
thermal occupation factors; it is distinct from the state-independent
coupling spectral density. Separate positive fits at each width would discard
the relevant physical constraint, as would separately normalizing away the
two filtered variance scales.

For known boxes of widths $d$ and $2d$, with $0<d<1/2$, the identity
\begin{equation}
 \frac{G_{2d}(\omega)^2}{G_d(\omega)^2}
   =\cos^2(\omega d/2),\qquad
 \frac{v_{2d}}{v_d}=\frac{1+x_{g,d}(d)}2
 \label{eq:widthidentity}
\end{equation}
shows that a second-width variance supplies an intermediate-time real
correlation. Once the sufficient integer moments and $v_d$ are matched,
Theorem~\ref{thm:clock} forces every thermal alias to move positive mass
from a hotter target frequency to its larger reflection. The first ratio
in equation~\eqref{eq:widthidentity} decreases strictly throughout the first
alias band, so every such alias has a smaller $v_{2d}$ than the target.
One added wide-width collective-variance measurement therefore removes these
ideal aliases after the required base-width moments and variance are fixed.
Changing the pulse width supplies information that the fixed-filter clock lacks.

We compare the two routes with the same target, collective force port, fresh preparations, loss and readout caps, matched references and counted control durations. Both pulse widths share a fixed ramp $r\in[.008,.012]$ and a common force area. An unknown common area factor is absorbed into $\mathcal V$, whose effective bare variance is bounded by the same collective-variance check.

\begin{figure}[!htbp]
\centering\includegraphics[width=0.85\linewidth]{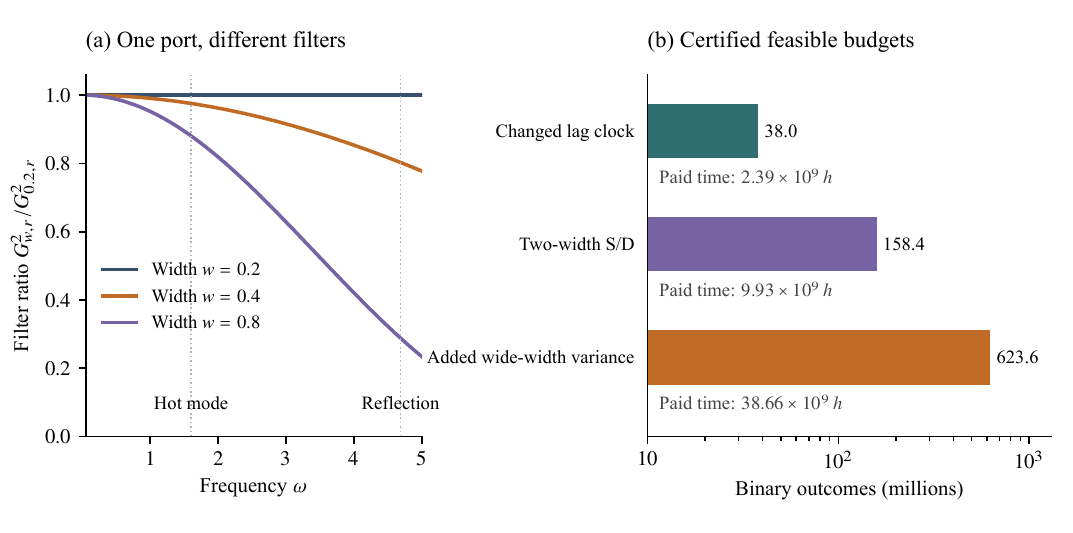}
\caption{Two measurement choices through the same collective port. (a) At the illustrative common ramp $r=.01$, the filter ratio decreases with frequency. Moving hot-mode mass from $1.6$ to its reflection $2\pi-1.6$ reduces the wider-pulse variance and exposes the integer-clock alias. The certificates cover the full ramp interval. (b) Sufficient acquisition budgets for the three designs in Table~\ref{tab:widthcost}. Bars count binary outcomes; annotations give acquisition-time upper bounds including matched references and fresh preparations. Specification and timing or width validation are additional.}
\label{fig:width}
\end{figure}

The selected design uses widths $.2,.8$ at integer lags $1,\ldots,4$. It requires about $4.17$ times as many binary outcomes as the fractional-clock design (Table~\ref{tab:widthcost} and figure~\ref{fig:width}). Adding only the wider-pulse variance needs fewer settings but substantially more outcomes. A tunable width therefore provides a certified option when the lag clock is fixed; changing the clock has the smaller acquisition budget in this comparison family.

These budgets guarantee at most $.05$ conditional type-I error and at least $.95$ power on a common alternative class. That class is the intersection of the clock neighborhood in equation~\eqref{eq:powerclass} with the corresponding joint-width neighborhoods. The original clock observations alone do not constrain the added coordinates. Outward bounds place the entire specified ramp/area target family in this intersection. All routes use the same tail and residual allowances, preparation-origin loss bounds, simultaneous confidence bounds, allocation procedure and 10\% count safety factor.

The width certificate enforces one spectrum, temperature and ramp across both widths, including the variance ratio $v_w/v_{.2}$. Its support is bounded at the temperature/ramp endpoints and over continuous frequency. Of eleven width/lag recipes, all have a geometric gap, but six fail the systematic-error floor of this sufficient test even with unlimited counts. Supplement S9 gives those recipes, exact directions, gates and allocations. The factor $4.17$ compares counted outcomes; specification and route-specific validation are additional resources.

\begin{table}[tbp]
\centering\small
\begin{tabular}{p{.27\linewidth}p{.24\linewidth}rr}
\toprule
Acquisition & Lags and widths & $C_{\rm probe}/10^6$ & Time bound$/10^9h$\\
\midrule
Fractional clock & $(.5,1,2,3,4)$; $.2$ & $38.011$ & $2.393$\\
Two-width $S/D$ & $(1,2,3,4)$; $.2,.8$ & $158.405$ & $9.927$\\
Added wide-width variance & $(1,2,3,4)$; $.2$, variance $.8$ & $623.644$ & $38.657$\\
\bottomrule
\end{tabular}
\caption{Selected certified sufficient budgets within the same witness and confidence-bound
family, including the 80,000-slot collective-variance stage and every matched
reference and preparation. Time uses the primitive upper bounds in
equation~\eqref{eq:time}; $h$ is a sampling unit. Every route additionally
requires $T_{\rm spec}$ and its failure allocation. The fractional-clock route
requires timing/phase validation $T_{\rm clock}$; width routes require
cross-width port, area, ramp, phase and heating validation $T_{\rm width}$.
These are acquisition guarantees, not complete experiment costs or minimum
sample complexities.}
\label{tab:widthcost}
\end{table}

\FloatBarrier

Finally, a tighter temperature interval does not automatically improve another task. For the one-lag nonfolding regime identified in Supplement S11, every positive variance spectrum fitting the fluctuation can also fit the measured response by choosing a common temperature. Response then leaves the compatible range of later pure-dephasing signals unchanged. This example shows why improved thermal discrimination must be linked to the observable required by the downstream task.

\section{Discussion}

An unresolved positive spectrum can trade mode occupation against coupling weight, making a non-equilibrium preparation appear thermal on an entire sampling lattice. The quantum response restricts this trade: positivity fixes the original and reflected frequency pairs, and a common temperature must be no warmer than the coldest target mode. This produces the temperature-order-dependent cutoff. A fractional-time observation resolves the reflected pairs; a second calibrated pulse width resolves them through the change in filtered variance.

For characterization through one collective port, the practical question is therefore which new observation to acquire. Our finite-error comparison includes the same reset, reference and control costs for both choices. Changing the clock has the smaller certified budget in the specified family, while changing width remains informative when the lag clock is fixed. Mode-resolved thermometry is substantially cheaper when individual addressing and mode frequencies are available.

The laboratory requirements are reproducible initial states, known spectral support and validated controls. Random interleaving accommodates drift when the two possible force choices share the same background within each slot. Independent controls must bound force-dependent heating, phase shifts and detector changes. The collective-variance stage then permits predictable changes in Gaussian loss between preparations.

The costs are sufficient acquisition guarantees for a specified harmonic module; they do not include commissioning or route-specific timing and width validation. Their useful parameter range is tabulated in the supplement, and their laboratory duration scales with the chosen sampling interval. Extending the analysis to several folding bands or nonlinear and non-Gaussian dynamics would require additional spectral constraints or observables. Within the present model, the result links a sharp equilibrium-identifiability boundary to an explicit choice of sampling time or pulse filter.

\section*{Data availability}
The data and code supporting this study are supplied as supplementary
data. They include exact witnesses, continuous-support certificate records,
positive thermal constructions, saved binary-count diagnostics, a fixed-input
manifest, software requirements and a verification entry point for the reported
numerical results.
\section*{Acknowledgments}
This work was supported by the Science, Technology and Innovation
Commission of Shenzhen Municipality (Grant Nos. JCYJ20170412152620376 and
KYTDPT20181011104202253), the Innovation Program for Quantum Science and
Technology (2021ZD0301703), Guangdong Major Project of Basic Research
(2025B0303000007), Quantum Science and Technology-National Science and
Technology Major Project (Grant No. 2024ZD0300400), and the Shenzhen Science
and Technology Program (Grant No. KQTD20200820113010023).
\section*{Author contributions}
\textbf{Zi-Xuan Wang:} Formal analysis (initial analytical derivations).
\textbf{Xiu-Hao Deng:} Conceptualization (overall research framework and
scientific narrative); Formal analysis (generalization of the analytical
results); Investigation (numerical calculations); Supervision.
\section*{Conflict of interest}
The authors declare no competing interests.



\clearpage

\renewcommand{\thesection}{S\arabic{section}}
\renewcommand{\theequation}{S\arabic{equation}}
\renewcommand{\thefigure}{S\arabic{figure}}
\renewcommand{\thetable}{S\arabic{table}}
\setcounter{section}{0}
\setcounter{equation}{0}
\setcounter{figure}{0}
\setcounter{table}{0}

\begin{center}
{\Large\bfseries Supplementary Material}

\smallskip
{\small Proofs, continuous-temperature certificates, finite-sample protocol details
and matched pulse-width comparisons for ``Thermal aliases in quantum noise:
sampling clocks and pulse widths.''}
\end{center}

\paragraph{Reading guide.}
For the model and identifiability proof, read S1--S3. For the physical controls and measured decision rule, read S5--S7; S4 derives the Gaussian-loss kernel used there. To recompute the selected budgets, use S9.4--S9.7 and their exact inputs; S8 supplies the diagnostic implementation details. S10 compares the old integer clock and mode-resolved thermometry, and S11 gives the task-transfer boundary.

\paragraph{Symbols and terminology.}
\begin{center}\small
\begin{tabular}{p{.24\linewidth}p{.66\linewidth}}
\toprule
Symbol or term & Meaning in the physical test\\\midrule
$q$, $v\,\dd q$ & Normalized mode-variance weights and the corresponding bare variance spectrum, including thermal occupation.\\[2pt]
$q_g$ & Probability measure of the pulse-filtered variance spectrum.\\[2pt]
$v_g$ & Total filtered collective variance; $v_g\,\dd q_g=G_d^2v\,\dd q$.\\[2pt]
$m_{\rm filt}$ & Known lower bound on squared filter magnitude over the band and allowed pulse shapes.\\[2pt]
$a_j$, $d_j$, $T_j$ & Score normalization weights, pulse widths and dimensionless coherent-window durations, respectively.\\[2pt]
$x_g$, $y_g$, $Y$ & Filtered fluctuation, response and their vector at the selected times.\\[2pt]
$z$ (witness) & Coefficients of a linear combination that can separate the target observations from thermal spectra.\\[2pt]
$H$ (support bound) & A certified upper bound on the thermal value of that combination on a specified temperature interval.\\[2pt]
Leaf & One interval in the cover of $u=\beta/(1+\beta)$; the width calculation also bounds the shared ramp.\\[2pt]
$u_j,t_j,p_j$; $p_A$ & Fluctuation, variance-scale and response statistics defined in the main text; $p_A$ is instead the force-on sampling probability.\\[2pt]
Score and phase branch & $F_z$ compares data with $H$. The phase-branch check selects the response without a $2\pi$ ambiguity.\\[2pt]
Gate; residual cap & A measured validity check; a supplied bound on unmatched complex-mean errors. Either can limit use of the test.\\[2pt]
Placebo; fixed horizon & An independent control that bounds force-label-dependent mismatch; the stream length chosen before acquisition.\\
\bottomrule
\end{tabular}
\end{center}

\section{Canonical model and finite-box influence functional}

The conventions are $[Q_k,P_l]=\ii\delta_{kl}$ and $a_k=(Q_k+\ii P_k)/\sqrt2$.
For a centered thermal mode with initial occupation $n_k$, put
\begin{equation}
 v_k=g_k^2(n_k+1/2),\qquad \gam_k=\frac{1}{2n_k+1}
   =\tanh(\beta_k\omega_k/2).
\end{equation}
Then $\nu(t,s)=\sum_kv_k\cos\omega_k(t-s)$ and
$\mu(t,s)=-\sum_kv_k\gam_k\sin\omega_k(t-s)$ are respectively the symmetrized
kernel and the commutator divided by $2\ii$. A fixed diagonal normalization
$v=\sum_kv_k$ gives the main-text kernel. These real and imaginary coordinates
come from the same Hamiltonian frequencies and positive coupling measure.

For the implemented force histories $-\lambda[h_1(t)|1\rangle\langle1|+h_0(t)|0\rangle\langle0|]B(t)$,
put $\Delta=h_1-h_0$ and $\Sigma=h_1+h_0$. Gaussianity and the scalar commutator give
\begin{equation}
 \log W=-\frac{\lambda^2}{2}\iint\Delta(t)\Delta(s)\nu(t,s)\dd t\dd s
 -\ii\lambda^2\int_{t>s}\Delta(t)\Sigma(s)\mu(t,s)\dd t\dd s.
 \label{eq:influence}
\end{equation}
The phase depends on both difference and sum histories. Replacing $XP_1X=P_0=I-P_1$
by $-P_1$ drops the unconditional force and is incorrect. Reversing the reported
sensor quadrature conjugates all phases consistently; changing only the inference
model's response sign is not a convention change.

Let $e=g_d(t)$, $l=g_d(t-\tau)$ with $g_d=1/d$ on $[0,d]$, $\tau>d$.
For $S$, $(\Delta,\Sigma)=(e+l,e+l)$; for $D$, $(\Delta,\Sigma)=(e-l,e+l)$.
The latter uses a population flip between boxes and a final restoration. The Fourier box
filter is $e^{\ii\omega d/2}G_d(\omega)$, $G_d=\sinc(\omega d/2)$. Define
\begin{align}
 v_g&=\sum_kv_kG_d(\omega_k)^2,\nonumber\\
 \nu_g(\tau)&=\sum_kv_kG_d(\omega_k)^2\cos\omega_k\tau,\nonumber\\
 R_g(\tau)&=\sum_kv_k\gam_kG_d(\omega_k)^2\sin\omega_k\tau,\nonumber\\
 S_d&=\sum_kv_k\gam_k f_d(\omega_k),\quad
 f_d(\omega)=\frac{\omega d-\sin\omega d}{(\omega d)^2}.
 \label{eq:self}
\end{align}
The ordered integral in $S$ adds both equal self phases and the cross phase;
in $D$ it subtracts the late self phase and the cross phase. Consequently
\begin{align}
 W_S&=e^{-\lambda^2(v_g+\nu_g)}e^{\ii\lambda^2(2S_d+R_g)},\nonumber\\
 W_D&=e^{-\lambda^2(v_g-\nu_g)}e^{-\ii\lambda^2R_g}.
\end{align}
Only the $D$ phase is used. A stable control phase $c\lambda^2$ per occupied
box contributes $2c\lambda^2$ in $S$ and zero in $D$. If early and late phases
differ, their difference does not cancel; it belongs in the residual mean bound.
A force-off phase reference alone cannot determine an unknown per-force-box
phase. A symmetric opposite-branch force has $\Sigma=0$ and no corresponding
ordered phase. The state-selective force retains the nonzero sum history
needed by this readout, including its unconditional contribution.

If $G_d^2>0$ on a compact band, every positive normalized $q_g$ has a physical
bare pullback. To match a selected $v_g$, an atom of filtered weight $p_l$ has
bare variance $v_l=v_gp_l/G_d(\omega_l)^2$ and coupling
$g_l^2=2v_l\gam_\beta(\omega_l)$. The total bare variance may change during this
refit. An externally known exact bare variance is therefore additional information;
it cannot be asserted when constructing an alias solely at fixed $v_g$.

An experimental oscillator rotating frame requires particular care. The tested
temperature is defined with respect to the Hamiltonian whose $\omega_k$ appears
in the time evolution and Boltzmann factors. A Gibbs state of a programmed
effective Hamiltonian can be the target of a noise emulator. Its fitted inverse
temperature is not automatically the laboratory temperature obtained by inserting
a microwave carrier frequency into the same formula.

\section{Proof of the finite-mode thermal cutoff}

Let $M\ge2$, $0<\omega_1<\cdots<\omega_M<\pi$, $p_i>0$, $\sum p_i=1$, and
finite positive $\beta_i$ not all equal. A null is a positive probability measure
on $[a,b]\subset(0,2\pi)$, containing the target frequencies, at one common
$\beta$. Put $c_i=\cos\omega_i$, $\Omega_i=2\pi-\omega_i$ and
$\beta_*=\max_i\beta_i$.

The product $Q(\omega)=\prod_i(\cos\omega-c_i)^2$ has a finite expansion in
$1,\cos\omega,\ldots,\cos(2M\omega)$. Its nonnegativity and zero target
expectation imply that any positive measure matching those moments is supported
only on its roots. Within $(0,2\pi)$ these are exactly $\omega_i$ and
$\Omega_i$, when present in the band. Write their total mass as $m_i$.
The first $M$ real equations, indexed $j=0,\ldots,M-1$, read
$\sum_i m_iT_j(c_i)=\sum_i p_iT_j(c_i)$. The polynomials $T_j$ have nonzero
leading coefficients and form a basis through degree $M-1$, so their evaluation
matrix is nonsingular at distinct $c_i$. More explicitly its determinant is
\begin{equation}
 2^{(M-1)(M-2)/2}\prod_{i<k}(c_k-c_i)\ne0.
\end{equation}
Hence $m_i=p_i$.

Likewise $\sin(j\omega_i)=\sin\omega_i\,U_{j-1}(c_i)$ for $j=1,\ldots,M$.
The determinant of this response matrix is
\begin{equation}
 2^{M(M-1)/2}\left(\prod_i\sin\omega_i\right)
                  \prod_{i<k}(c_k-c_i)\ne0.
\end{equation}
The response data therefore identify each pair coefficient, giving
\begin{equation}
 p_i\gam_{\beta_i}(\omega_i)
 =(p_i-s_i)\gam_\beta(\omega_i)-s_i\gam_\beta(\Omega_i),\quad 0\le s_i\le p_i.
 \label{eq:pair}
\end{equation}
Its right side is at most $p_i\gam_\beta(\omega_i)$, so strict monotonicity in
$\beta$ forces $\beta\ge\beta_*$. Every $i$ with $\beta_i<\beta_*$ needs
$s_i>0$. Thus each of their reflections must fit in the band, proving the
necessity of $b\ge b_c=\max_{\beta_i<\beta_*}\Omega_i$.

For sufficiency choose $\beta=\beta_*$. For a coldest mode set $s_i=0$; for
each hotter mode set
\begin{equation}
 s_i=p_i\frac{\gam_{\beta_*}(\omega_i)-\gam_{\beta_i}(\omega_i)}
                  {\gam_{\beta_*}(\omega_i)+\gam_{\beta_*}(\Omega_i)}.
\end{equation}
This lies strictly between zero and $p_i$ and solves equation~\eqref{eq:pair}.
The cosine and sine coefficients then match for every integer $j$, not only
the finite equations used in the necessity proof. Positive physical couplings
follow from S1. At least one coldest mode is unsplit, giving at most $2M-1$
atoms. Ties among the coldest temperatures simply reduce their number.

For an added $0<r<1$, matching the real observation requires
\begin{equation}
 0=\sum_i s_i[\cos(r\omega_i)-\cos(r\Omega_i)]
  =2\sin(\pi r)\sum_i s_i\sin[r(\pi-\omega_i)].
\end{equation}
All factors multiplying $s_i$ are positive. Thus all reflected weights vanish,
and equation~\eqref{eq:pair} requires $\beta=\beta_i$ for every mode, a contradiction.
At $\beta=0$ every pair response is zero; after reflections vanish the target
responses are positive in the coefficient basis. At $\beta=\infty$ every
unreflected pair coefficient is $p_i$, exceeding its target value. Below $b_c$
at least one required pair is unreflected, which excludes that endpoint as well.

The temperature compactification $u=\beta/(1+\beta)$ is continuous on $[0,1]$
when $a>0$, including the uniform $\beta\to\infty$ limit of $\gam_\beta$.
Probability measures on a compact band are weakly compact. Their joint finite
moment image over $u$ is therefore compact. Exclusion of exact equality proves
strictly positive distance to this closed set. It gives neither a uniform margin
over all target parameters nor a numerical value from a finite dictionary.

The theorem applies unchanged to positive filtered weights. Its assumptions are positive weights, distinct target frequencies strictly between $0$ and $\pi$, a finite target mode count, and one folding band. If all temperatures coincide the target is already a null. As $r\to0$
the added cosine difference can scale as $r^2$, and as $r\to1$ it vanishes;
small temperature contrast or coalescing frequencies can also collapse the gap.

For the two-mode example at width $0.2$, the filtered weights are
approximately $0.50172675$ and $0.49827325$. The common-temperature alias
at $\beta=1.7$ uses $0.7$, $1.6$ and $4.683185\ldots$ as its frequencies.
It matches $v_g=0.994931736991\ldots$ and has initial bare variance about
$1.01149$, below the coarse cap $2$. In the unfiltered construction, the
three weights are $0.5$, $0.32915746$ and $0.17084254$. Pulse filtering
changes the weights, so each construction uses its own normalization. Independent three-mode construction checks also include
two weight distributions, different temperature orderings and a true common
temperature; they provide arithmetic consistency checks, while the argument above proves the theorem.

\section{All-temperature continuous support certificates}

At a fixed $\beta$, positive measures form the convex hull of the atom map.
For $z=(z_R,z_I)$ define
\begin{equation}
 C_z(\omega)=\sum_jz_{R,j}\cos\omega\tau_j,\qquad
 S_z(\omega)=\sum_jz_{I,j}\sin\omega\tau_j.
\end{equation}
For each fixed $\omega$, $C_z+\gam_\beta S_z$ is monotone in $\beta$ in one
direction or the other. Taking the supremum over a temperature interval and
then frequency proves the exact endpoint identity in the main text. The sign
of $S_z$ need not be fixed across the band. This identity concerns a support
upper bound, not a physical mixture of endpoint temperatures.

At fixed $\beta$, the derivative satisfies
\begin{equation}
 |\partial_\omega\gam_\beta(\omega)|
 =\frac\beta2\operatorname{sech}^2(\beta\omega/2)
 \le\min\{\beta/2,1/a\}.
\end{equation}
For the second bound write $x=\beta\omega/2$ and use
$x\operatorname{sech}^2x\le1$. For example,
$\cosh^2x=1+\sinh^2x\ge1+x^2\ge x$ for $x\ge0$.
This gives the stated Lipschitz constant. At the compactified endpoints
$\gam=0$ or $1$ is constant, so this derivative contribution vanishes.
Every point lies within $\Delta\omega/2$ of a grid node. A node maximum plus
$L\Delta\omega/2$ is therefore an upper bound on the continuous support.

Each temperature leaf stores its exact rational endpoints in $u$, rational
$z$ with exact $\ell_1$ norm at most one, interval target coordinates, and
outward node/support bounds at both endpoint temperatures. If
$\delta_l=z_l\cdot Y-H_l>0$ for every leaf covering $[0,1]$, the minimum is a
lower bound on all-temperature sup-norm distance. An observed component error
$r$ subtracts at most $r\|z_l\|_1$. A physical upper bound uses one common
temperature and actual positive atomic weights; a convex combination of
different-temperature fits is not accepted as such an upper bound.

If a target has positive distance from the entire compact thermal image, local
separating witnesses exist at every temperature and persist in an open
neighborhood by continuity. A finite subcover exists. Finding that cover is a separate numerical task, and a finite search may remain inconclusive.

For the narrow-band point study, the discovery dictionary has 1001 frequencies
and the final support grid 2049 nodes. The two-, three- and four-lag covers have
4, 6 and 11 leaves, respectively. All 86,058 stored endpoint nodes were read
back independently at higher precision. For the band-$5$, width-$0.2$ clock
comparison, each positive schedule has two leaves, $u\in[0,1/2]$ and
$u\in[1/2,1]$, corresponding to $\beta\in[0,1]$ and $[1,\infty]$.
The final support grids have 8193 nodes per endpoint; 65,544 node values were
independently checked. Both programs use the same interval arithmetic library,
so the independence is of expressions and implementations, not of arithmetic
engines. Exact theory supplies the between-node and temperature coverage.

If an initial bare tail has mass at most $\ell$, the filtered in-band mass is
at least $m(1-\ell)$ and the unnormalized filtered tail is at most $\ell$.
Thus $\ell_g\le\ell/[m(1-\ell)+\ell]$. Every atom has support at most
$\|z\|_1$, while the band support is at most both $H_{\rm band}$ and that norm.
For any actual tail mass $t\le\ell_g$,
\begin{equation}
 H_{\rm full}\le(1-t)\min(H_{\rm band},\|z\|_1)+t\|z\|_1
 \le(1-\ell_g)\min(H_{\rm band},\|z\|_1)+\ell_g\|z\|_1.
\end{equation}
At $d=0.2$, band $[0.05,5]$ and $\ell=0.001$, $m=0.919395388264\ldots$ and
$\ell_g<0.001087576$. One outward-rounded rational value of $H$ is used
throughout all score and error coefficients.

\subsection{Numerical certificates for the reported distance bounds}
The coordinates are ordered as all real moments followed by all positive response moments, $-\operatorname{Im}k$. Write $u=\beta/(1+\beta)$, and let $H$ be the larger certified endpoint support on a leaf. The unfiltered maps $I_2,I_3,I_4$ use $(1,2)$, $(1,2,3)$ and $(1,2,3,4)$ on $[1/20,5/2]$. The box-filtered maps $B_{\rm ins}$ and $B_{1/2}$ use $(1/2,1,2,3,4)$ and $(1/2,1,3/2,2)$ on $[1/20,5]$, with box width $d=1/5$. Their target weights include the normalized positive box-filter power.

These five lower bounds concern different observation maps and do not order their exact distances. The three upper constructions are for the three unfiltered maps only.
\FloatBarrier
\begin{table}[!htbp]
\centering\scriptsize
\setlength{\tabcolsep}{4pt}
\caption{All 25 temperature leaves of the five continuous-frequency certificates. Target-dot lower bounds and margins round downward, while support upper bounds round upward. The closures include $\beta=0$ and $\beta=\infty$.}
\label{tab:revision-leaf-cover}
\begin{tabular}{@{}lccrrr@{}}\hline
Map:leaf & $u$ interval & $\beta$ interval & $(z\cdot Y)_{\rm L}$ & $H_{\rm U}$ & $\delta_{\rm L}$\\\hline
\multicolumn{6}{l}{$I_2$: unfiltered normalized moments; all-$\beta$ lower bound $0.0216294$}\\
$I_2$:000 & $[0,\tfrac{1}{4}]$ & $[0,\tfrac{1}{3}]$ & $0.2229509$ & $0.0963735$ & $0.1265774$\\
$I_2$:001 & $[\tfrac{1}{4},\tfrac{3}{8}]$ & $[\tfrac{1}{3},\tfrac{3}{5}]$ & $0.0285322$ & $-0.0080186$ & $0.0365509$\\
$I_2$:002 & $[\tfrac{3}{8},\tfrac{1}{2}]$ & $[\tfrac{3}{5},1]$ & $-0.1207380$ & $-0.1423673$ & $0.0216294$\\
$I_2$:003 & $[\tfrac{1}{2},1]$ & $[1,\infty]$ & $-0.1910290$ & $-0.2841141$ & $0.0930852$\\
\hline
\multicolumn{6}{l}{$I_3$: unfiltered normalized moments; all-$\beta$ lower bound $0.0985292$}\\
$I_3$:000 & $[0,\tfrac{1}{4}]$ & $[0,\tfrac{1}{3}]$ & $0.2216628$ & $0.0620971$ & $0.1595657$\\
$I_3$:001 & $[\tfrac{1}{4},\tfrac{3}{8}]$ & $[\tfrac{1}{3},\tfrac{3}{5}]$ & $0.1827287$ & $0.0648124$ & $0.1179164$\\
$I_3$:002 & $[\tfrac{3}{8},\tfrac{7}{16}]$ & $[\tfrac{3}{5},\tfrac{7}{9}]$ & $0.1670519$ & $0.0604282$ & $0.1066237$\\
$I_3$:003 & $[\tfrac{7}{16},\tfrac{15}{32}]$ & $[\tfrac{7}{9},\tfrac{15}{17}]$ & $0.1627933$ & $0.0589625$ & $0.1038309$\\
$I_3$:004 & $[\tfrac{15}{32},\tfrac{1}{2}]$ & $[\tfrac{15}{17},1]$ & $0.0299747$ & $-0.0685545$ & $0.0985292$\\
$I_3$:005 & $[\tfrac{1}{2},1]$ & $[1,\infty]$ & $-0.0629303$ & $-0.1753996$ & $0.1124694$\\
\hline
\multicolumn{6}{l}{$I_4$: unfiltered normalized moments; all-$\beta$ lower bound $0.1324644$}\\
$I_4$:000 & $[0,\tfrac{1}{4}]$ & $[0,\tfrac{1}{3}]$ & $0.2260439$ & $0.0643131$ & $0.1617309$\\
$I_4$:001 & $[\tfrac{1}{4},\tfrac{5}{16}]$ & $[\tfrac{1}{3},\tfrac{5}{11}]$ & $0.1904744$ & $0.0470108$ & $0.1434637$\\
$I_4$:002 & $[\tfrac{5}{16},\tfrac{3}{8}]$ & $[\tfrac{5}{11},\tfrac{3}{5}]$ & $0.1581101$ & $0.0256458$ & $0.1324644$\\
$I_4$:003 & $[\tfrac{3}{8},\tfrac{13}{32}]$ & $[\tfrac{3}{5},\tfrac{13}{19}]$ & $0.1591599$ & $0.0209185$ & $0.1382415$\\
$I_4$:004 & $[\tfrac{13}{32},\tfrac{7}{16}]$ & $[\tfrac{13}{19},\tfrac{7}{9}]$ & $0.1606706$ & $0.0230639$ & $0.1376067$\\
$I_4$:005 & $[\tfrac{7}{16},\tfrac{15}{32}]$ & $[\tfrac{7}{9},\tfrac{15}{17}]$ & $0.1622922$ & $0.0255880$ & $0.1367043$\\
$I_4$:006 & $[\tfrac{15}{32},\tfrac{1}{2}]$ & $[\tfrac{15}{17},1]$ & $0.1640319$ & $0.0285788$ & $0.1354532$\\
$I_4$:007 & $[\tfrac{1}{2},\tfrac{9}{16}]$ & $[1,\tfrac{9}{7}]$ & $0.0445147$ & $-0.0881444$ & $0.1326592$\\
$I_4$:008 & $[\tfrac{9}{16},\tfrac{5}{8}]$ & $[\tfrac{9}{7},\tfrac{5}{3}]$ & $0.0044139$ & $-0.1565056$ & $0.1609196$\\
$I_4$:009 & $[\tfrac{5}{8},\tfrac{3}{4}]$ & $[\tfrac{5}{3},3]$ & $-0.0729589$ & $-0.2801973$ & $0.2072384$\\
$I_4$:010 & $[\tfrac{3}{4},1]$ & $[3,\infty]$ & $-0.0402897$ & $-0.3305008$ & $0.2902111$\\
\hline
\multicolumn{6}{l}{$B_{\rm ins}$: normalized box filtered; all-$\beta$ lower bound $0.0649854$}\\
$B_{\rm ins}$:000 & $[0,\tfrac{1}{2}]$ & $[0,1]$ & $0.1854624$ & $0.1204771$ & $0.0649854$\\
$B_{\rm ins}$:001 & $[\tfrac{1}{2},1]$ & $[1,\infty]$ & $-0.0258044$ & $-0.1229181$ & $0.0971137$\\
\hline
\multicolumn{6}{l}{$B_{1/2}$: normalized box filtered; all-$\beta$ lower bound $0.0020988$}\\
$B_{1/2}$:000 & $[0,\tfrac{1}{2}]$ & $[0,1]$ & $0.1238239$ & $0.1217252$ & $0.0020988$\\
$B_{1/2}$:001 & $[\tfrac{1}{2},1]$ & $[1,\infty]$ & $-0.1377095$ & $-0.2307826$ & $0.0930731$\\
\hline
\end{tabular}
\end{table}
\begin{table}[!htbp]
\centering\footnotesize
\setlength{\tabcolsep}{3pt}
\caption{Exact rational witnesses on the leaf attaining each map's smallest certified margin. Each continued list gives all coordinates of $A=Dz$ in order, and the norm is exact. Displayed decimals do not replace the rational values used in the certificates.}
\label{tab:revision-rational-witnesses}
\begin{tabular}{@{}l>{\raggedright\arraybackslash}p{0.21\linewidth}>{\raggedright\arraybackslash}p{0.67\linewidth}@{}}\hline
Map:leaf & Denominator and norm & Exact integer components $A=Dz$\\\hline
$I_2$:002 & $D=50000000000000000$\newline $\|z\|_1=\tfrac{49999999999499997}{50000000000000000}$ & \(\begin{gathered}-12904107614429122,913566601363891,-18650550947778599,\\ 17531774835928385\end{gathered}\)\\[4pt]
$I_3$:004 & $D=100000000000000000$\newline $\|z\|_1=\tfrac{19999999999800001}{20000000000000000}$ & \(\begin{gathered}0,-22512955006024515,11690412802369828,\\ -33785599312683351,15935256841087395,16075776036834916\end{gathered}\)\\[4pt]
$I_4$:002 & $D=5000000000000000000$\newline $\|z\|_1=\tfrac{4999999999950000259}{5000000000000000000}$ & \(\begin{gathered}0,-594570862085240850,455007849819366805,\\ 259376700150610595,0,1510424458201490550,\\ 2147730229473059650,32889900220231809\end{gathered}\)\\[4pt]
$B_{\rm ins}$:000 & $D=10^{20}$\newline $\|z\|_1=\tfrac{99999999999000007279}{10^{20}}$ & \(\begin{gathered}45252797704605559,0,-1340397145365939500,\\ 714290076198564620,10576250819996842000,0,\\ 28603503828347343000,20885395577243296000,\\ 28619709426055256000,9215200328088160600\end{gathered}\)\\[4pt]
$B_{1/2}$:000 & $D=1000000000000000000$\newline $\|z\|_1=\tfrac{999999999989999959}{1000000000000000000}$ & \(\begin{gathered}0,338306628430938890,-346625818320520080,\\ 128629461756827410,65720973885923684,0,\\ -55362831853356768,65354285742433127\end{gathered}\)\\[4pt]
\hline\end{tabular}
\end{table}
\begin{table}[!htbp]
\centering\footnotesize
\setlength{\tabcolsep}{4pt}
\caption{Three positive-spectrum upper constructions for the unfiltered maps. Frequencies and normalized weights are exact rationals, and all atoms in a construction have its displayed common inverse temperature. The final residual column rounds the outward $\ell_\infty$ upper bound upward.}
\label{tab:revision-physical-uppers}
\begin{tabular}{@{}lcllr@{}}\hline
Map & Common $\beta$ & Frequency & Positive weight & Residual upper\\\hline
$I_2$ & $0.7071520500984703$ & $\dfrac{20943}{20000}$ & $\dfrac{27233378107953371}{500000000002740259}$ & $0.0274257$\\
$I_2$ &  & $\dfrac{656}{625}$ & $\dfrac{433204746322463585}{500000000002740259}$ & \\
$I_2$ &  & $\dfrac{5}{2}$ & $\dfrac{39561875572323303}{500000000002740259}$ & \\
\hline
$I_3$ & $0.9545431885259403$ & $\dfrac{16239}{20000}$ & $\dfrac{54670437314208220}{99999999999999753}$ & $0.1057936$\\
$I_3$ &  & $\dfrac{509}{625}$ & $\dfrac{14396022670005923}{99999999999999753}$ & \\
$I_3$ &  & $\dfrac{36427}{20000}$ & $\dfrac{6130197872035657}{33333333333333251}$ & \\
$I_3$ &  & $\dfrac{9119}{5000}$ & $\dfrac{4180982133226213}{33333333333333251}$ & \\
\hline
$I_4$ & $1.0841318361859356$ & $\dfrac{3337}{5000}$ & $\dfrac{11361681105662773}{100000003512002861}$ & $0.1467426$\\
$I_4$ &  & $\dfrac{13397}{20000}$ & $\dfrac{36910094008099081}{100000003512002861}$ & \\
$I_4$ &  & $\dfrac{6031}{4000}$ & $\dfrac{31034594032890700}{100000003512002861}$ & \\
$I_4$ &  & $\dfrac{7551}{5000}$ & $\dfrac{10134141877355556}{100000003512002861}$ & \\
$I_4$ &  & $\dfrac{5}{2}$ & $\dfrac{10559492487994751}{100000003512002861}$ & \\
\hline
\end{tabular}
\end{table}
\FloatBarrier
\subsection{A physical bracket for the broad-band distance comparison}
\label{supp:distanceorder}

Let $\delta_4$ denote the true all-temperature sup-norm distance of the
width-$1/5$ box-filtered target at lags $(1/2,1,3/2,2)$ on $[.05,5]$,
and $\delta_5$ the distance at $(1/2,1,2,3,4)$. The lower bars alone do
not order these distances. The following additional positive construction
at one common inverse temperature $\beta=13/20$ supplies the additional upper bound needed.
Its filtered weights sum to one exactly.

\begin{center}
\small
\begin{tabular}{cc}
\toprule
Exact frequency & Exact positive filtered weight\\
\midrule
$43529/40960$ & $5608771211574623/99999999999998741$\\
$10907/10240$ & $3778218367844830/4347826086956467$\\
$13423/5120$ & $3704029783609406/99999999999998741$\\
$107483/40960$ & $3788176544383622/99999999999998741$\\
\bottomrule
\end{tabular}
\end{center}

For any chosen $v_g>0$, set
\begin{equation}
 v_k=\frac{v_g p_k}{G_d(\omega_k)^2},\quad
 n_k=\frac{1}{e^{\beta\omega_k}-1},\quad
 g_{Q,k}^2=2v_k\tanh(\beta\omega_k/2).
 \label{eq:distancephysical}
\end{equation}
These are finite positive bare oscillator parameters in the manuscript
convention $B=\sum g_{Q,k}Q_k$, $Q_k=(a_k+a_k^\dagger)/\sqrt2$.
The same construction can instead be written as
$B=\sum g_{a,k}(a_k+a_k^\dagger)$, with $g_{Q,k}=\sqrt2g_{a,k}$.
At $v_g=1$ the total bare variance is
$1.00524738997\ldots$.

Interval evaluation of the actual oscillator correlations gives
\begin{equation}
 .00209880\le\delta_4\le .04050609
 < .0649854\le\delta_5.
 \label{eq:distanceorder}
\end{equation}
The final lower bound is the continuous-frequency/all-temperature certificate
already tabulated above. The upper bound was found in a fixed finite
dictionary of 24 rational temperatures and 2049 frequencies; every row was
evaluated, and the dictionary is not a restriction of the thermal null.
The proof of the upper uses the exact positive atoms, not an optimizer
objective or a failed search. A normalized cosine/thermal-factor calculation
at 100 digits, a separate Bose/box-integral reconstruction at 110 digits,
and a primary $Q$-operator reconstruction at 120 digits enclose the same
vector. All share the interval arithmetic library, so arithmetic-engine
independence is not asserted.

Equation~\eqref{eq:distanceorder} orders true distances for two different
observation maps. It establishes neither optimal distance nor a preparation
or time advantage; those questions need their own readout and cost design.
\FloatBarrier

\section{Nonstationary Gaussian loss and its observation bound}

For one initially thermal mode, let $a=\kappa/2$, $v=g^2(n_0+1/2)$,
$\gam=1/(2n_0+1)$ and loss-bath occupation $0\le n_b\le1/2$.
The time origin for these kernels is the first force, at which the repeated
thermal state is specified. Section S6 treats the finite-control module whose
preparation precedes the initial Ramsey, shifts the force histories from
preparation completion and includes that additional loss interval explicitly.
The parameters $\kappa,n_b$ are fixed within a campaign, as are the repeated
closed-state preparation and pulse envelope. Uniform bounds permit any
admissible fixed choice, not arbitrary shot-to-shot changes of the loss map.
Its linear Langevin solution or Gaussian unitary dilation gives, for $t\ge s\ge0$,
\begin{align}
 \nu_\kappa(t,s)&=g^2\left[(n_b+1/2)e^{-a(t-s)}
                    +(n_0-n_b)e^{-a(t+s)}\right]\cos\omega(t-s),\nonumber\\
 \mu_\kappa(t,s)&=-\frac{g^2}{2}e^{-a(t-s)}\sin\omega(t-s).
 \label{eq:losskernels}
\end{align}
The $t+s$ contribution is essential when the initial state relaxes toward a
different reservoir occupation. A stationary damped cosine omits it.
Gaussianity of the full dilation makes equation~\eqref{eq:influence} valid during
the force pulse with these kernels. Number dephasing $\mathcal D[a^\dagger a]$
or Kerr evolution has no such general Gaussian-dilation reduction.

Since $2\gam|n_b-n_0|\le1$, writing the symmetric kernel as
$v e^{-a(t-s)}+g^2(n_b-n_0)e^{-a(t-s)}(1-e^{-2as})$ times the cosine yields
\begin{equation}
 |\nu_\kappa-\nu_0|\le va(t+s),\qquad
 |\mu_\kappa-\mu_0|\le v\gam a|t-s|.
 \label{eq:kernelbounds}
\end{equation}
For a larger known cap $0\le n_b\le N$, define
$C_N=\max\{1,2N\}$. Since
$|n_b-n_0|/(n_0+1/2)\le C_N$, the same decomposition gives
$|\nu_\kappa-\nu_0|\le C_Nva(t+s)$; the commutator bound is unchanged.
The cap $N=1/2$ gives the unit coefficient used in the reported budgets.
For a warmer reservoir the symmetric loss allowance must be enlarged and
the finite-count guarantees re-evaluated.

For both $S,D$ sequences, $|\Delta|=\Sigma=e+l$, with integral $2$ and first moment $\tau+d$.
Consequently the symmetric logarithm error is at most
$2\lambda^2va(\tau+d)$. The ordered integral of $|\Delta(t)\Sigma(s)|(t-s)$ is
$\tau+d/3$: its cross part is $\tau$ and each self box contributes $d/6$.
The phase error is at most $\lambda^2v\gam a(\tau+d/3)$.
Summing modes and using $\gam\le1$, $v_{\rm total}\le V_{\max}$ and
$a\le a_{\max}$ gives the force-start complex-logarithm bound. Section S6 includes the initial Ramsey interval measured from bath preparation.

Both the coherent and lossy logarithms have nonpositive real parts. Integrate
the exponential along their connecting line segment to obtain
$|e^{L_1}-e^{L_0}|\le|L_1-L_0|$. This proves the complex-mean error bound,
uniformly in frequency, temperature and oscillator Hilbert-space dimension.
Multiplication by a sensor visibility no greater than one cannot increase it.

For reproducibility, the exact integrals also have simple closed forms. Write
$z=a-\ii\omega$ and
\begin{equation}
 A(z)=\frac{zd-1+e^{-zd}}{(zd)^2},\quad
 F(z)=\frac{1-e^{-zd}}{zd},\quad
 B(z,\tau)=e^{-z\tau}\left[\frac{\sinh(zd/2)}{zd/2}\right]^2.
\end{equation}
Put $v_b=g^2(n_b+1/2)$ and $v_\Delta=v-v_b$. Use $s=+1$ for $S$ and $s=-1$ for $D$.
The real parts and measured phases are
\begin{align}
 \Re\log W_s&=-\lambda^2\left[v_b(2\Re A+s\Re B)
 +\frac{v_\Delta}{2}|F|^2|1+s e^{-z\tau}|^2\right],\nonumber\\
 \Im\log W_S&=\lambda^2v\gam(2\Im A+\Im B),\qquad
 \Im\log W_D=-\lambda^2v\gam\Im B.
 \label{eq:exactloss}
\end{align}
The commutator depends only on $t-s$, even though the covariance is nonstationary.
Thus the translated self integrals in $D$ cancel at nonzero loss. The first-box
variance is $V_\kappa=2v_b\Re A+v_\Delta|F|^2$, and its phase is
$\lambda_0^2v\gam\Im A$. Independent quadratures of the two kernels and
selected Lindblad calculations at two oscillator truncations check the exact
map. Finite truncation is a consistency control, not a proof of the dimension-free bound.

Independent sensor Markov relaxation/dephasing multiplies the off-diagonal block by a common complex scalar; relaxation jumps change populations without feeding that coherence. Each signal and reference must match in complex phase and magnitude, including the population-flip controls. The matched reference phases can differ between $S$ and $D$; only the $D$ phase enters the score.

For a fixed binary POVM write $M=2E-I=m_0I+m_xX+m_yY+m_zZ$.
The operational order is evolution, final analysis rotation $R_{q,\pm}$,
independent fair $I/Z$, then detector. Averaging the coin gives
$(M+ZMZ)/2=m_0I+m_zZ$ for every sensor state. Trusted rotations satisfying
$R_{q,\pm}^\dagger ZR_{q,\pm}=\pm q$, $q=X,Y$, and opposite-phase balancing
give the same unknown $m_z$ on both quadratures and cancel $m_0$.
The common sign of $m_z$ cancels in each complex signal/reference ratio.
Very small gain causes inconclusive modulus/collective-variance gates. No detector tomography
is required under this fixed-POVM/trusted-control model. The coin is random for
each physical trial; conditioning on a favorable coin imbalance would change
the sampling law. Nonstationary detector/control errors are not removed and
must be bounded relative to the same closed comparison mean in every batch.

\section{Final pulse, control and validation specification}
\label{supp:joint}

This section collects the physical premises used by both selected designs.
They define one mathematical harmonic module. A provider must implement or
validate them jointly; the thermal-test outcomes do not infer them.

\subsection{Positive finite-ramp pulses}

Let $u_L$ be the unit-area uniform density on $[0,L]$ and let
$f_r=u_{d-r}*u_r$, where $d=.2$ and $.008\le r\le.012$.
This positive trapezoid has unit area, support $[0,d]$ and centered transform
\begin{equation}
 G_r(\omega)=\sinc[\omega(d-r)/2]\sinc(\omega r/2).
 \label{eq:trapcontrol}
\end{equation}
On $[.05,5]$, both sinc factors are positive and
$G_r(\omega)\ge G_d(\omega)>0$. For fixed $\omega$,
\begin{equation}
 \partial_r\log G_r=(\omega/2)
 \{f'[\omega r/2]-f'[\omega(d-r)/2]\}\ge0,
 \qquad f(t)=\log\sinc t,
\end{equation}
because $f'$ decreases on this interval. The minimum over frequency and ramp
is therefore attained at $(\omega,r)=(5,.008)$ and can be enclosed without a
ramp grid. A common force-area factor $A\in[.9999,1.0001]$ multiplies all bare
variances by $A^2$ and cancels from normalized weights. Area, ramp and the
collective coordinate must be common to every history and, for the two-width
route, common across both widths.

Independent outward evaluation over the two target modes gives
\begin{align}
 .9951208698674&<v_g<.9957010654860,\nonumber\\
 0<&v_g-v_{g,0}<.000769329,\nonumber\\
 \|Y-Y_0\|_\infty&<.000495670,\qquad v\le1.00020001<1.01.
 \label{eq:rampdomain}
\end{align}
These bounds cover the continuous ramp and area family and place it inside the
power domain used for the final count calculation.

For a pulse-time draw, write
$T=(1-r/d)U+(r/d)V$ with independent uniform $U,V$ on $[0,d]$.
For an independent copy $T'$,
\begin{equation}
 \mathbb E|T-T'|\le(1-r/d)\mathbb E|U-U'|
 +(r/d)\mathbb E|V-V'|=d/3.
\end{equation}
Together with unit area and equal pulse centers, this retains the loss bounds
used in section S4. Additional smoothing or transfer-function error must fit
the residual envelope or be recertified.

\subsection{Finite gates, visibility and residual allocation}

The finite-control example uses
\begin{equation}
 t_X=.05h,\quad t_R=t_Z=.01h,\quad
 t_{\rm settle}\le.02h,
\end{equation}
with the force off during rotations. The final $Z$ slot includes its fair coin
and conditional-control latency. Including the analysis-label choice gives a
conservative coherent overhead $.18h$. The shortest $.3h$ free gap accommodates
the population flip and settling interval.

For detector contrast at least $.9$ and Markovian sensor dephasing
$T_2\ge10^4h$, every selected window obeys
\begin{equation}
 .9\exp[-(w+.18)/10^4]>.9\exp(-w/8).
 \label{eq:completevisibility}
\end{equation}
A total deterministic unitary-channel error at most $2\times10^{-5}$, together
with the dephasing comparison, gives a combined complex-mean contribution below
$10^{-4}$. The remaining $9\times10^{-4}$ envelope covers calibrated
preparation, pulse and matched-reference errors; individual sources must share
this allowance rather than each receiving it in full.

Within-slot signal/reference matching is an independently validated premise.
A dummy-waveform, detuned or zero-coupling placebo sequence must bound
force-label-dependent phase, heating and detector changes without using the
unknown spectrum or temperature. Its bound enters the same complex-mean
envelope. The validation record must state its confidence, refresh rule and
validity interval. A failed or expired validation returns inconclusive.

\subsection{Route-specific validation and cost boundary}

\begin{center}
\begin{tabular}{p{.20\linewidth}p{.45\linewidth}p{.22\linewidth}}
\toprule
Route & Additional validation & Cost outside reported acquisition\\\midrule
Both & Band/tail, harmonicity, fresh preparation, loss caps, detector twirl and within-slot matching & $T_{\rm spec},\alpha_{\rm spec}$\\[2pt]
Fractional clock & Lag and analysis-phase timing at the inserted half step & $T_{\rm clock},\alpha_{\rm clock}$\\[2pt]
Two widths & Same port, area and ramp; width-dependent transfer function, phase, heating and cross-talk & $T_{\rm width},\alpha_{\rm width}$\\
\bottomrule
\end{tabular}
\end{center}

The acquisition totals in the article count every signal/reference outcome,
fresh bath preparation, force box, swap, Ramsey and final randomized-$Z$
primitive. They do not assign the validation entries above a zero cost. Thus
the reported factor between the two selected routes orders acquisition outcomes
under a common specification; it is not a complete hardware-cost ordering.

For a conditional scale illustration, $h=1\,\mu\mathrm s$ maps the counted
times to about 40 minutes, 2.76 hours and 10.74 hours for the three displayed
routes before $T_{\rm spec}$, $T_{\rm clock}$ and $T_{\rm width}$. It also
requires 10 ns Ramsey rotations, 50 ns population flips, 8--12 ns ramps and
$T_2\ge10$ ms simultaneously. These numbers set the requirements for a possible
realization. Their joint implementation requires platform validation.
\section{Collective-variance gate for predictable inter-slot loss}
\label{supp:variableloss}

Time zero is completion of bath preparation immediately before the initial
Ramsey rotation. The first force begins at
$0\le t_{\rm pre}\le.01h$. In slot $i$, every
$\kappa_{k,i},n_{b,k,i}$ may depend predictably on earlier records, provided it
is chosen before the fresh force label, remains constant through that coherent
slot and is shared by its two counterfactual force choices. The closed state,
spectrum, pulse and reset procedure remain fixed. Within-slot loss modulation,
changed preparations and force-label-dependent heating are outside this class.

\subsection{A noncircular observable bound}

For a binary POVM, $-I\le M=2E-I\le I$. After the correctly ordered twirl, the
ideal reference coherence is $B_i=m_{z,i}c_i e^{\ii\phi_i}$ and satisfies
$|B_i|\le1$. This bounds the conditional mean without estimating detector gain.
Shifting both force boxes by $t_{\rm pre}$ changes the covariance moment while leaving the stationary commutator unchanged. Let $m_{\rm filt}$ be a supplied lower bound on squared filter magnitude throughout the band and pulse family. For the trapezoids of S5, $G_{d,r}\ge G_d>0$, so we use $m_{\rm filt}=\min_{\omega\in[.05,5]}G_d^2(\omega)=\sinc^2(.5)\simeq.919395$. The pulse and loss bounds give
\begin{equation}
 V_{E,i}\ge K_{\min}v,\qquad
 K_{\min}=m_{\rm filt}(1-\ell)
 -\bar\kappa(d+2t_{\rm pre})/2>0,
 \qquad |W_{E,i}|\le e^{-K_{\min}v/2}.
 \label{eq:per_slot_energy}
\end{equation}
The collective-variance compensator need not factor, but obeys
\begin{equation}
 A_E^*=\frac1{n_E}\sum_iB_iW_{E,i},\qquad
 |A_E^*|\le e^{-K_{\min}v/2}.
\end{equation}
On the simultaneous coverage event define
\begin{equation}
 a_L=|\widehat A_E|-e_E,\qquad
 v_{U,\rm abs}=-\frac{2\log a_L}{K_{\min}}.
 \label{eq:absoluteenergy}
\end{equation}
Continue only if $0<a_L\le1$ and $v_{U,\rm abs}\le2$; otherwise report
inconclusive. A passing gate implies $v\le v_{U,\rm abs}$ and never rejects by
itself. No reference ratio or phase cancellation enters this implication.

After the gate, the uniform two-box signal allowance is
\begin{equation}
 \epsilon_{{\rm loss},j}^{\rm pre}(2)
 =2\lambda^2\frac{\bar\kappa}{2}
 \sqrt{4(\tau_j+d+2t_{\rm pre})^2+(\tau_j+d/3)^2}.
\end{equation}
Writing $W_{q,0}$ for the closed comparison factor,
\begin{equation}
 A_q^*=W_{q,0}\overline B_q+
 \frac1{n_j}\sum_iB_{q,i}(W_{q,i}-W_{q,0}),
\end{equation}
the second term is bounded by this allowance in the signal disk. The residual
cap is not enlarged to hide loss. The martingale, logarithmic-disk, phase and
thermal-support arguments are otherwise unchanged, so the pathwise rejection
proof has conditional type-I error at most $.05$ for this enlarged null class.

\subsection{Uniform passage on the stated alternative class}

For the single-box stage,
\begin{equation}
 |W_{E,i}-W_{E,0}|\le\epsilon_{E,\Gamma}^{\rm pre}
 =1.01\,\frac{\bar\kappa}{2}
 \sqrt{\frac{(d+2t_{\rm pre})^2}{4}+\frac{d^2}{36}}.
\end{equation}
With $n_E=80{,}000$, put
\begin{equation}
 a_\Gamma=\eta_E e^{-v_{g,U}/2}
 -\epsilon_{E,\Gamma}^{\rm pre}-2e_E.
\end{equation}
Outward evaluation gives
\begin{center}
\begin{tabular}{ll}\toprule
Quantity & Conservative bound\\\midrule
$K_{\min}$ & $>.9184671928$\\
$\epsilon_{E,\Gamma}^{\rm pre}$ & $<4.644\times10^{-6}$\\
$e_E$ & $<.038195$\\
$a_\Gamma$ & $>.4570908979$\\
$-2\log a_\Gamma/K_{\min}$ & $<1.704739<2$\\
\bottomrule\end{tabular}
\end{center}
Thus the collective-variance gate passes throughout the stated coverage event.
Re-evaluating the downstream guards at the selected safe counts gives positive
leaf and phase-branch margins. These margins, rather than the finite diagnostic
datasets, establish power at least $.95$ uniformly over the stated alternative,
loss and nonvanishing-reference domain.
\section{Randomized matched acquisition and rejection rule}
\label{supp:random}

The final design uses a horizon fixed before acquisition. Each slot begins
from the same prescribed closed Gaussian preparation and pulse family. Loss
parameters may vary predictably between fresh slots as specified in section
S6; detector/Ramsey factors and bounded residual means may also vary with the
past. The within-slot signal/reference background remains common and is
validated independently as described in section S5.

\subsection{Conditional means and a common random compensator}

For a fixed protocol, lag and width, let $\mathcal F_{i-1}$ contain the past
before slot $i$. Draw a fresh force label $I_i$ with probability $p$ and a
fresh analysis label uniformly from four phases. After the correctly ordered
fair detector twirl, suppose
\begin{align}
 \mathbb E[s_i\mid I_i=1,\theta_i,\mathcal F_{i-1}]
 &=o_i+\Re(e^{-\ii\theta_i}C_{A,i}),\nonumber\\
 \mathbb E[s_i\mid I_i=0,\theta_i,\mathcal F_{i-1}]
 &=o_i+\Re(e^{-\ii\theta_i}C_{B,i}),\nonumber\\
 C_{A,i}&=B_iW_i+\delta_{A,i},\qquad
 C_{B,i}=B_i+\delta_{B,i}.
 \label{eq:conditional-law}
\end{align}
Here $B_i,o_i$ may be predictable functions of previous observations,
$|B_i|\le1$, and the residual conditional means are bounded. The same
counterfactual factor $B_i$ must apply to both possible fresh force labels.

For
\begin{equation}
 X_{A,i}=\frac{2I_is_i e^{\ii\theta_i}}p,\qquad
 X_{B,i}=\frac{2(1-I_i)s_i e^{\ii\theta_i}}{1-p},
\end{equation}
the four-phase identities give
\begin{equation}
 \mathbb E[X_{A,i}\mid\mathcal F_{i-1}]=C_{A,i},\qquad
 \mathbb E[X_{B,i}\mid\mathcal F_{i-1}]=C_{B,i}.
\end{equation}
Thus $\widehat A=n^{-1}\sum_iX_{A,i}$ and
$\widehat B=n^{-1}\sum_iX_{B,i}$ estimate matched predictable averages even
when $\overline B=n^{-1}\sum_iB_i$ is random. The proof neither treats
$\widehat A/\widehat B$ as unbiased nor conditions on a future reference.

\subsection{Fixed-horizon confidence disks}

For either scalar component $X$ of $X_A$, $|X|\le2/p$ and
$\mathbb E[X^2\mid\mathcal F_{i-1}]=2/p$. The centered increment therefore
satisfies $|Z|\le4/p$ and conditional variance at most $2/p$.
The Bernstein--martingale inequality\cite{Freedman1975} gives the complex
radius
\begin{equation}
 e(n,p,L)=\sqrt{\frac{8L}{pn}}+\frac{8\sqrt2L}{3pn}.
 \label{eq:suppmartingale}
\end{equation}
For $N$ retained complex streams, use
$L_A=\log(4N/\alpha_A)$ and $L_B=\log(4N/\alpha_B)$ with the prescribed
horizon of each stream. The separate collective-variance stream uses
$p_E=1/2$ and $L_E=\log(8/\alpha_E)$. Residual compensators enter through
the triangle inequality, while $|B_i|\le1$ transfers the Gaussian loss bound
to each signal disk. With
$(\alpha_A,\alpha_B,\alpha_E)=(.02,.02,.01)$, simultaneous disk failure is
at most $.05$. Physical-specification failure is additional unless its
probability is allocated and the counts are recertified in advance.

\subsection{Observable gates and all-temperature rejection}

The collective-variance gate of section S6 first bounds every passing null by
$v\le2$. For every retained complex mean require $|\widehat X|>e_X$ and set
\begin{equation}
 d_X=-\log(1-e_X/|\widehat X|).
\end{equation}
This disk controls both log modulus and local phase. For witness $z$, support
$H$ and fixed nonnegative weights $a_j$, define
\begin{align}
 c_{D,j}&=\sqrt{(z_{R,j}+Ha_j)^2+4z_{I,j}^2},\nonumber\\
 c_{S,j}&=|-z_{R,j}+Ha_j|,\nonumber\\
 E_z&=\sum_j[c_{D,j}(d_{AD,j}+d_{BD,j})
             +c_{S,j}(d_{AS,j}+d_{BS,j})].
\end{align}
After the variance gate, $|\lambda^2R_w|\le2\lambda^2<\pi$. The measured
$D$ phase therefore has the correct branch whenever
\begin{equation}
 d_{AD,j}+d_{BD,j}<\pi-2\lambda^2
\end{equation}
for every retained $D$ stream. A coefficient that vanishes in all temperature
leaves is omitted before acquisition.

The bare-tail allowance is
$B_z=2\lambda^2V_{\max}\ell(\|z\|_1+|H|)$ with $V_{\max}=2$.
Reject only if the variance, nonzero-modulus and phase gates pass and the raw
score satisfies $\widehat F_z>B_z+E_z$ on both temperature intervals.
Otherwise report inconclusive. On the simultaneous coverage event a true
common-temperature positive variance spectrum belongs to one covered leaf and
cannot pass this rejection inequality. This proves conditional type-I error at
most $.05$ over the full continuous frequency and temperature domains.

The power calculation uses prespecified lower moduli, the moment/variance
neighborhood of the target and the same double-error disk bounds. Counts are
increased until every leaf has positive outward-rounded margin, then multiplied
by the fixed 1.1 safety factor. Section S9 prints the resulting separate-stream
horizons for the fractional-clock and pulse-width designs.

The limitations are observable in the protocol. Four equally weighted
reference phases $0,\pi/2,\pi,3\pi/2$ give a vanishing average reference and
therefore an inconclusive gate. A force-label-dependent factor such as
$B_A=.9$, $B_B=.9\ii$ is not repaired by randomization and fails the matching
contract. Neither case is reported as evidence against thermal equilibrium.
\section{Raw-count interface and numerical diagnostics}
\label{supp:randomcounts}

The numerical diagnostic uses the certified trapezoid with $d=.2$, $r=.01$,
unit area, $\kappa h=10^{-6}$ and $n_b=0$. Let
$A=d-r$, $B=r$ and $z=\kappa/2-\ii\omega$. Its exact transforms are
\begin{align}
 F(z)&=\frac{1-e^{-zA}}{zA}\frac{1-e^{-zB}}{zB},\nonumber\\
 J(z)&=\int_0^d C_f(u)e^{-zu}\dd u,\nonumber\\
 C_f(u)&=\frac{1}{6A^2B^2}
 \sum_{k,l=0}^2(-1)^{k+l}\binom2k\binom2l
 [d+u-kA-lB]_+^3 .
 \label{eq:trapkernels}
\end{align}
An independent calculation propagates the causal convolution
$K'+zK=f$ through the three linear segments of $f$ and then integrates $fK$.
The two formulas agree below the prespecified high-precision tolerance at all
checked frequency/rate points. Uniform guarantees still come from the analytic
and outward interval bounds rather than these point diagnostics.

For one mode, writing $a=\kappa/2$ and
$X_\tau=e^{-z\tau}F(z)F(-z)$ gives
\begin{align}
 V_e&=g^2[(n_b+\tfrac12)2\Re J+(n_0-n_b)|F(z)|^2],\nonumber\\
 V_l&=g^2[(n_b+\tfrac12)2\Re J+(n_0-n_b)e^{-2a\tau}|F(z)|^2],\nonumber\\
 C_\tau&=g^2\Re[(n_b+\tfrac12)X_\tau
 +(n_0-n_b)e^{-z\tau}|F(z)|^2],\nonumber\\
 S&=(g^2/2)\Im J,\qquad R=(g^2/2)\Im X_\tau.
\end{align}
After summing modes,
\begin{align}
 W_S&=\exp[-\lambda^2(V_e+V_l+2C_\tau)/2
 +\ii\lambda^2(2S+R)],\nonumber\\
 W_D&=\exp[-\lambda^2(V_e+V_l-2C_\tau)/2-\ii\lambda^2R],\nonumber\\
 W_E&=\exp[-V_e/2+\ii S].
 \label{eq:trapnative}
\end{align}
The occupation-dependent covariance is kept nonstationary; only the commutator
is stationary, which is enough to cancel the two $D$ self phases.

For each prespecified stream with horizon $n_j$, record nonnegative integer
counts with shape $[2,4,2,2]$: force label, analysis phase, twirl coin and
binary outcome. All 32 cells must sum to $n_j$. If $D_{A,\theta}$ is the
signed outcome count summed over the twirl coin, then
\begin{equation}
 \widehat A_j=\frac{2}{p_A n_j}
 [D_{A,0}-D_{A,\pi}+\ii(D_{A,\pi/2}-D_{A,3\pi/2})],
\end{equation}
with the analogous expression for the reference and $1-p_A$.
The implementation rejects invalid shapes or totals, negative or noninteger
counts and truth fields. Temperature, spectrum, detector gain, reference phase
and generator probabilities never enter the decision API.

Diagnostics cover the target, common-temperature controls, the physical
reflected alias, a weak-contrast alternative, a vanishing-reference case and
an intentionally mismatched force-label case. The target records reject and
the thermal controls remain inconclusive under the final rule; the vanishing
reference fails an observable gate. The deliberately mismatched case is marked
outside the contract regardless of its numerical decision. An independently
written reader reproduces every stored final decision. These finite records
test the implementation and data interface; the outward deterministic margins
supply the stated uniform guarantees.
\section{A matched same-port pulse-width baseline}
\label{supp:width}

This section extends the available controls by allowing a second force
width while preserving the fixed-filter theorem's scope. The comparison uses
the preparation-origin, predictably changing Gaussian-loss class of section S6.
It uses the bandwidth, bare-tail, reset and control specifications summarized in the main-text protocol table.

\subsection{Shared physical spectrum and an exact width identity}

Write $\dd\mathcal V=v\dd q$ for the initial unfiltered
fluctuation-variance measure. Under the thermal null,
$\dd\mathcal R=\gam_\beta\dd\mathcal V$ is its response measure.
For an atom of variance mass $v_k$, the physical coupling is
$g_k^2=2v_k\gam_\beta(\omega_k)$ in the convention of S1.
The measure $\mathcal V$ therefore contains the occupation factor and is
not a state-independent coupling spectrum. Each thermal-null hypothesis has one
common $\beta$, one common $\mathcal V$ and one fixed common waveform
parameter, used at both widths and all lags. The spectrum and temperature
may refit together between hypotheses. A common unknown force-area
factor $A$ is absorbed into $\mathcal V$ as $A^2$; the collective-variance gate bounds
this same effective bare variance. An independent area for each width
would require a different null and is not covered here.

For known box waveforms, take $0<d<1/2$ and the S2 target with
$0<\omega_1<\cdots<\omega_M<\pi$, positive masses and unequal finite
temperatures. Suppose a common-temperature spectrum on
$[a,b]\subset(0,2\pi)$ matches $v_d$, the first $2M$ integer cosine
moments and the first $M$ integer response moments at width $d$.
S2 applied to the positive filtered measure fixes the original/reflected
pairs and their total masses. Every hotter target mode has reflected mass
$s_i>0$. Since
\begin{equation}
 A_{2d}(\omega)=\frac{G_{2d}^2}{G_d^2}
      =\cos^2(\omega d/2),\qquad
 A_{2d}'(\omega)=-\frac d2\sin(\omega d)<0,
 \quad 0<\omega<2\pi,
 \label{eq:widthexact}
\end{equation}
the two filtered variances differ by
\begin{equation}
 v_{2d}^{\rm target}-v_{2d}^{\rm null}
 =v_d\sum_i s_i
       [A_{2d}(\omega_i)-A_{2d}(2\pi-\omega_i)]>0.
 \label{eq:widthaliasgap}
\end{equation}
The $\beta=0$ limit already fails the nonzero response, and the
$\beta=\infty$ limit still needs positive reflected mass.
Thus adding $v_{2d}$ removes every such alias in the entire first alias
band. The joint finite vector has a positive distance from the compact
all-temperature closure, although this argument does not give its size.
An equivalent identity is
$v_{2d}/v_d=[1+x_{g,d}(d)]/2$.
This observation is obtained by a new pulse setting, not by reprocessing
the original fixed-filter integer data. For the displayed $\beta=1.7$
box alias, $v_{.4}^{\rm target}-v_{.4}^{\rm alias}=.0302136\ldots$;
that particular difference is not the minimum over all thermal nulls.

\subsection{Joint finite-ramp geometry and continuous certificates}

For widths $d=.2$ and $w>d$, use the positive unit-area pulses
$u_{d-r}*u_r$ and $u_{w-r}*u_r$, with one fixed common
$r\in[.008,.012]$. Their real filters are
\begin{equation}
 G_{w,r}=\sinc[(w-r)\omega/2]\sinc(r\omega/2),\qquad
 A_r(\omega)=\frac{G_{w,r}^2}{G_{d,r}^2}
 =\left\{\frac{\sinc[(w-r)\omega/2]}
                {\sinc[(d-r)\omega/2]}\right\}^2.
 \label{eq:rampratio}
\end{equation}
The common ramp factor cancels only because the two pulses share $r$.
Let $\dd q_d=G_{d,r}^2\dd\mathcal V/v_d$ on the band.
For a list of width/lag pairs $(d_j,\tau_j)$, define
\begin{equation}
 \Phi_{\beta,r}(\omega)=
 \big((A_j\cos\omega\tau_j)_j,
      (A_j\gam_\beta\sin\omega\tau_j)_j,A_r\big),
 \quad A_j=\begin{cases}1&d_j=d,\\ A_r&d_j=w.\end{cases}
 \label{eq:jointwidthatom}
\end{equation}
Its last coordinate is $v_w/v_d$. Every joint observation vector belongs
to $\operatorname{conv}\Phi_{\beta,r}([a,b])$ for one common
temperature and ramp. Allowing arbitrary $q_d$ covers its dependence on
the unknown bare measure and ramp; no fixed mode count is assumed.

On $[.05,5]$, $w\in\{.4,.6,.8\}$ and the stated ramp interval,
all sinc arguments remain in $(0,\pi)$. With
$\psi(x)=\cot x-1/x<0$ and $\psi'(x)<0$, differentiation gives
\begin{align}
 \partial_\omega\log A_r
   &=(w-r)\psi[(w-r)\omega/2]-(d-r)\psi[(d-r)\omega/2]<0,\nonumber\\
 \partial_r\log A_r
   &=\omega\{\psi[(d-r)\omega/2]-\psi[(w-r)\omega/2]\}>0.
 \label{eq:ratiomonotone}
\end{align}
In particular $0<A_r\le1$. A known identical ramp under both hypotheses
gives the same exact pair obstruction as equation~\eqref{eq:widthaliasgap}.
When a competing model can instead choose a different allowed common
$r$, that monotonicity argument alone is insufficient. The following
certificates cover this additional uncertainty for the specified target.

At fixed $\omega$, a witness score has the form
$a+b\gam_\beta+A_r(c+e\gam_\beta)$. It is affine separately in
$\gam_\beta$ and $A_r$, hence is bounded by its four corner values at
the temperature and ramp endpoints. The interpolation weights may depend
on frequency; they do not define a physical mixture of temperatures.
Two leaves, $u=\beta/(1+\beta)\in[0,1/2]$ and $[1/2,1]$, cover
$\beta\in[0,\infty]$. Each has its own rational direction $z$ with
exact $\|z\|_1\le1$.

For $m=.9193<\min G_{d,r}^2$, the derivative bound
$L_A=(w+d)/(2m)$ follows from $A_r\le1$ and
$|(G_{w,r}^2)'|\le w/2$. A valid support Lipschitz bound is
\begin{equation}
 L=\sum_j |z_{C,j}|(L_{A,j}+\tau_j)
  +|z_{{\rm resp},j}|(L_{A,j}+\tau_j+c_\beta)
  +|z_V|L_A,
 \quad c_\beta=\min\{\beta/2,1/a\},
 \label{eq:widthlip}
\end{equation}
where $L_{A,j}=0$ at width $d$ and $L_A$ at width $w$;
$c_\beta=0$ for the two limiting temperatures. At each of the four
corners, 8193 rational frequency nodes with step $4.95/8192$ are evaluated
outward at 50 decimal digits. The largest node upper endpoint plus
$L\Delta\omega/2$ is a continuous support upper bound $H$.
Each exact node interval and its ramp/temperature label is archived.
The minimum of $z\cdot Y_0-H$ across leaves is a certified lower
bound on the band-limited all-temperature sup-norm distance, not an
exact distance or a finite-shot guarantee.

The target center has the original frequencies, bare variance masses
$(1/2,1/2)$ and mode temperatures, with $r=.01$.
Table~\ref{tab:widthrecipes} records every fixed recipe, including
all finite-error failures. The exact directions are supplied as rational
values with the numerical data; Table~\ref{tab:widthdirections} prints the
selected full two-width directions as terminating decimals, without
rounding. Support upper bounds and gap lower bounds are outward records.

\FloatBarrier
\begin{table}[!htbp]
\centering\small
\begin{tabular}{llrrr}
\toprule
Data at integer lags & $m$ & Second width & Gap lower & $C_{\rm probe}/10^6$\\
\midrule
Two-width $S/D$ & 2 & 0.4 & 0.00002345 & --\\
Two-width $S/D$ & 3 & 0.4 & 0.01224542 & --\\
Two-width $S/D$ & 4 & 0.4 & 0.01249594 & --\\
Two-width $S/D$ & 2 & 0.6 & 0.00098599 & --\\
Two-width $S/D$ & 3 & 0.6 & 0.03149395 & 2089.311\\
Two-width $S/D$ & 4 & 0.6 & 0.03284503 & 2222.592\\
Two-width $S/D$ & 2 & 0.8 & 0.00234028 & --\\
Two-width $S/D$ & 3 & 0.8 & 0.03756854 & 204.527\\
Two-width $S/D$ & 4 & 0.8 & 0.05318316 & 158.405\\
Base $S/D$ + wide variance & 4 & 0.4 & 0.01079769 & --\\
Base $S/D$ + wide variance & 4 & 0.8 & 0.03460898 & 623.644\\
\bottomrule\end{tabular}
\caption{All eleven prespecified same-port width recipes. Gaps are continuous-band, all-temperature lower bounds including the common ramp nuisance. Counts are upward-rounded summaries of the selected finite-error sufficient allocations, not optima. A dash denotes a negative infinite-count upper margin in this fixed proof family, not physical impossibility.}
\label{tab:widthrecipes}
\end{table}
\begin{table}[!htbp]
\centering\scriptsize
\setlength{\tabcolsep}{2pt}
\begin{tabular}{rrrrrr}
\toprule
$w$ & $\tau$ & $z_C$, leaf 1 & $z_{\rm resp}$, leaf 1 & $z_C$, leaf 2 & $z_{\rm resp}$, leaf 2\\
\midrule
0.2 & 1 & 0 & 0.16552529408258587 & 0 & 0\\
0.2 & 2 & -0.021750664922827484 & 0 & 0 & 0\\
0.2 & 3 & 0.0080128442758821659 & 0 & 0 & 0\\
0.2 & 4 & 0.10188055913055744 & 0.1331799032666883 & 0 & 0.030771571694873288\\
0.8 & 1 & 0 & 0 & 0 & -0.27742987783650863\\
0.8 & 2 & 0 & 0.37929926646194939 & -0.17978810552661897 & 0\\
0.8 & 3 & 0 & 0.19034827915876115 & 0.025754531601594265 & 0.11271553246083978\\
0.8 & 4 & 0.0000031886007374835003 & 0 & 0.021097981568641455 & 0\\
\midrule
\multicolumn{2}{l}{$z_V$} & \multicolumn{2}{r}{0} & \multicolumn{2}{r}{0.35244239921092346}\\
\multicolumn{2}{l}{Support $H$ upper} & \multicolumn{2}{r}{0.1251528326} & \multicolumn{2}{r}{0.2736009401}\\
\multicolumn{2}{l}{Nominal gap lower} & \multicolumn{2}{r}{0.0648056316} & \multicolumn{2}{r}{0.0531831699}\\
\bottomrule\end{tabular}
\caption{Exact terminating-decimal directions for the selected $.2/.8$ integer-lag geometry. Leaves are $\beta\in[0,1]$ and $[1,\infty]$. Support values are rounded upward and gaps downward. The rational support records and directions for all recipes are supplied with the numerical data.}
\label{tab:widthdirections}
\end{table}
\FloatBarrier

\subsection{Bare tail, raw score and observable gates}

The two-width filter quotient is used only on the supplied band. Let
$f_z$ be the corresponding unnormalized kernel, so the band certificate
is $f_z\le HG_{d,r}^2$. Outside the band, the unit-area positive pulses
give $|G_{w,r}|\le1$. Thus, following the absolute collective-variance gate $v\le2$,
\begin{equation}
 F_z=2s\int(f_z-HG_{d,r}^2)\dd\mathcal V
 \le B_z=2sV_{\max}\ell(\|z\|_1+|H|),
 \quad s=\lambda^2,\quad V_{\max}=2.
 \label{eq:widthtail}
\end{equation}
No division by a possibly zero base filter is made in the tail. The allowance includes one
shared bare-tail fraction $\ell=.001$. The same conservative
allowance is used for the matched clock comparison. The fixed-width clock
certificate has its distinct, tighter filtered-tail accounting.

For each pair use the main-text raw $u_j,t_j,p_j$. Predetermined
normalization weights $a_j$ sum to one over base-width pairs; $b_j$
sum to one over second-width pairs. Write $k_j=-Ha_j+z_Vb_j$.
The score expands directly as
\begin{align}
 F_z&=\sum_j[z_{C,j}u_j+z_{{\rm resp},j}p_j+k_jt_j],\nonumber\\
 c_{D,j}&=\sqrt{(z_{C,j}-k_j)^2+4z_{{\rm resp},j}^2},\qquad
 c_{S,j}=|-z_{C,j}-k_j|.
 \label{eq:widthraw}
\end{align}
These are the coefficient norms multiplying the $D$ complex log and
$S$ real log. The direct complex coefficient is
$(z_{C,j}-k_j)+2\ii z_{{\rm resp},j}$; the real/imaginary coordinate
vector is $(z_{C,j}-k_j,-2z_{{\rm resp},j})$.
For an added single-box wide-width variance with strength $s_E=1.7$,
$\log|C_E|=-s_Ev_w/2$ and its coefficient is $-4sz_V/s_E$.
This variance stream uses the main simultaneous confidence pools. The
separate 80,000-slot absolute collective-variance stage is still included once.

A stream is omitted only if its expanded coefficient is exactly zero
in both leaves and no retained gate needs it. This avoids constructing
an unmeasured $u_j$ or $t_j$: the expanded raw score is the actual
statistic. On simultaneous mean-disk coverage, define
$d_X=-\log(1-e_X/|\widehat X|)$ and require positive modulus disks.
After the absolute collective-variance gate, $|sR_w|\le2s<\pi$ for every used
positive waveform. The additional observable condition
\begin{equation}
 d_{A_D}+d_{B_D}<\pi-2s
 \label{eq:widthbranch}
\end{equation}
ensures that $\arg(\widehat A_D/\widehat B_D)$ has the correct
principal branch, even when the individual mean phases cross their cuts.
No $S$ observation enters this gate. A failed gate is inconclusive.
The branch argument and absolute-variance bound hold pathwise on the
coverage event; they do not condition concentration on a future reference.

With $E_z=\sum_Xc_X(d_{A_X}+d_{B_X})$, reject only when every leaf
satisfies $\widehat F_z>B_z+E_z$. A true null belongs to one covered
leaf, whose true score is at most $B_z$; it cannot reject on the
simultaneous coverage event. For the actual retained channel count $N$,
use $L_A=L_B=\log(4N/.02)$ in the section S7 martingale radius and retain
$\alpha_E=.01$. Predetermined unequal horizons and omission of zero
streams leave the martingale proof intact. Conditional type-I error
is at most $.02+.02+.01=.05$. Specification failure remains additional
unless the statistical allocation is changed in advance.

\subsection{Common power domain, allocation and acquisition costs}

For the comparison, intersect the old clock moment box with the new
joint-coordinate boxes $\|Y-Y_0\|_\infty\le.001$, retaining
$v_d\in[v_{g,0}-.001,v_{g,0}+.001]$, effective bare variance
$v\le1.01$ and the original reference floors
$\eta_{w,\tau}=.9e^{-(\tau+w)/8}$.
The additional coordinates are power premises, not consequences inferred
from the old data, and are not supplied as exact metadata to the null test.
The continuous common-area/ramp target family is nonempty in this
intersection: outward endpoint/Cartesian bounds give old-clock moment
deviation $<.000409251$, new-coordinate deviation $<.000448055$ for
every fixed recipe, variance deviation $<.000769329$ and
$v\le1.00020001$. Monotonicity encloses all ramps, not a sampled ramp
grid. The matched phase cone, pure-$Z$ dephasing and finite gates of section S5
also meet every added reference floor. At width $.8$, the shortest
integer force-off gap is $.2h$, exceeding the $.07h$ population-flip/settling
primitive; two ramps still fit the $.024h$ switching allowance.
The section S6 preparation clock and all predictable inter-slot loss caps are
used. Equal area and ramp are additional control specifications,
whose validation cost is charged separately.

For a retained mean with ideal modulus lower bound $R>2e$, score
displacement and the observed threshold together cost exactly
\begin{equation}
 L_X=\log\frac R{R-2e}
 =\log\frac R{R-e}+\log\frac{R-e}{R-2e}.
 \label{eq:widthdouble}
\end{equation}
The sum accounts for both displacement of the measured score and its observed error threshold.
At the base width the signal modulus uses one moment radius; at the
second width the variance-ratio and cosine radii add, giving $2(.001)$.
The preparation-origin loss allowance uses each actual width in the
section S6 positive-waveform bound. A sufficient leaf margin is
\begin{equation}
 2s v_{d,L}[\delta_z-.001\|z\|_1]-B_z
       -\sum_X c_X(L_{A_X}+L_{B_X})>0,
 \label{eq:widthpower}
\end{equation}
with the same collective-variance gate and a sufficient $D$ branch condition
$L_{A_D}+L_{B_D}<\pi-2s$. The absolute-variance power upper is
$<1.704760<2$, using a conservative variance cap 2 for its small
loss allowance.
Together these guards imply at least $.95$ power uniformly on the
declared domain; vanishing average references are still excluded from
the power statement.

The same finite search is run for all eleven width recipes and the clock:
$s\in\{.7,.9,1.1\}$, $p_A\in\{.5,.75\}$, one-hot normalization
over all admissible pairs, and deterministic counts per retained stream.
Using $y_X=\sqrt{10^7/n_X}$, SLSQP seeks a feasible preparation budget
with $10^4\le n_X\le2\times10^9$, at most 400 iterations and at
most four alternations with normalization enumeration. Counts are
inflated by 10\% and rounded upward, then verified independently of
the optimizer by 70-digit outward expressions. Counts, coefficients
and normalizations are fixed before outcomes; no outcome-dependent
choice is used. The certified output is a feasible allocation in the specified search family.
For each failure in Table~\ref{tab:widthrecipes}, an outward upper
bound on the best infinite-count leaf margin remains negative over
every allowed one-hot choice at each specified strength. The fixed
sufficient-proof family therefore fails there; other witnesses or error
models remain open. Ideal-box and uniform-count calculations provide
exploratory comparisons under different allocation restrictions.

\begin{table}[tbp]
\centering\small
\begin{tabular}{llrrr}
\toprule
Acquisition & Width & Lag & $n_S$ & $n_D$\\
\midrule
Fractional clock & 0.2 & 0.5 & 152,774 & 459,271\\
 & 0.2 & 1 & 0 & 7,264,979\\
 & 0.2 & 2 & 687,855 & 9,439,850\\
 & 0.2 & 3 & 561,119 & 11,312,156\\
 & 0.2 & 4 & 1,025,878 & 7,027,362\\
\midrule
Two-width $S/D$ & 0.2 & 1 & 19,645,037 & 14,544,208\\
 & 0.2 & 2 & 464,101 & 712,244\\
 & 0.2 & 3 & 293,567 & 369,854\\
 & 0.2 & 4 & 712,171 & 9,549,010\\
 & 0.8 & 1 & 23,452,987 & 25,411,121\\
 & 0.8 & 2 & 11,739,000 & 19,141,587\\
 & 0.8 & 3 & 3,765,410 & 20,591,538\\
 & 0.8 & 4 & 3,917,475 & 4,015,331\\
\midrule
Added wide-width variance & 0.2 & 1 & 0 & 66,045,913\\
 & 0.2 & 2 & 123,410,304 & 129,129,368\\
 & 0.2 & 3 & 23,574,253 & 62,160,152\\
 & 0.2 & 4 & 3,428,851 & 13,967,177\\
 & $.8$ variance & -- & \multicolumn{2}{r}{201,847,610}\\
\midrule
\multicolumn{3}{l}{Collective-variance gate, each route} & \multicolumn{2}{r}{80,000}\\
\bottomrule\end{tabular}
\caption{Prespecified trials per stream. Each number counts randomized force and matched reference slots together, using its prespecified denominator; realized force-label counts are not postselected. A zero denotes an exactly omitted score stream. The collective-variance horizon is additional to the listed main streams, once per route.}
\label{tab:widthcounts}
\end{table}
\begin{table}[tbp]
\centering\small
\begin{tabular}{lrrr}
\toprule
Counted operation & Fractional clock & Two-width $S/D$ & Added wide variance\\
\midrule
$C_{\rm probe}=C_{\rm bath}$ & 38,011,244 & 158,404,641 & 623,643,628\\
$C_{\rm box}$ & 75,942,488 & 316,729,282 & 1,045,359,646\\
$C_{\rm swap}$ & 71,007,236 & 188,669,786 & 542,605,220\\
$C_{\rm Ramsey}$ & 76,022,488 & 316,809,282 & 1,247,287,256\\
$T_{\rm window}/h$ & 103261445.3 & 393905869.6 & 1143749875.6\\
\bottomrule\end{tabular}
\caption{Exact resource totals. Twirl, force-label and analysis-label counts each equal $C_{\rm probe}$. All potential force boxes and reference preparations are counted. Specification, clock and width-validation costs are separate.}
\label{tab:widthprimitives}
\end{table}

The selected clock and full two-width designs both use $s=1.1$,
$p_A=.75$. Their minimum leaf margins exceed $.00469509$ and
$.00345503$, respectively; the added-wide-width-variance design has margin
$>.00142460$. Table~\ref{tab:widthcounts} gives their exact horizons.
The clock normalization uses lag 4 in the first leaf and lag 2 in the
second. The full two-width design uses base lags 4 and 1, respectively,
and second-width lag 1. The single-box variance recipe uses base lags 4 and 2.
The clock alone requires no new-coordinate premise to obtain its
38,011,244-outcome guarantee; its old power box suffices. Comparing
both routes uses the smaller common intersection described above.

Every slot pays a full probe/bath preparation and matched padded
force-off schedule. A two-box $S$ slot pays two boxes, a $D$ slot
also pays two population flips, and a single-box variance slot pays one box. Each pays
two Ramsey rotations and all label/twirl primitives. Let $C_X$ denote
their totals and $T_{\rm window}$ the sum of actual width/lag windows.
For the supplied primitive caps,
\begin{equation}
 T_{\rm paid}/h=T_{\rm window}/h+60.05C_{\rm probe}
              +.07C_{\rm swap}+.024C_{\rm box}.
 \label{eq:widthcost}
\end{equation}
The selected upper time bounds, including specification costs, are
\begin{align*}
 T_{\rm frac,total}&=2{,}392{,}629{,}773.732h+T_{\rm spec}+T_{\rm clock},\\
 T_{\rm two\text{-}width,total}&=9{,}926{,}912{,}949.438h+T_{\rm spec}+T_{\rm width},\\
 T_{\rm wide\ variance,total}&=38{,}656{,}620{,}733.904h+T_{\rm spec}+T_{\rm width}.
\end{align*}
$T_{\rm clock}$ and $T_{\rm width}$ are unpriced supplied validation costs,
not zero. These upper time bounds deliberately retain operations that overlap
an evolution window. Fewer settings or a larger geometric gap alone
does not imply a cheaper acquisition.

The comparison gives two certified feasible acquisition routes and a
conditional outcome-cost ordering within the stated witness, waveform and
confidence-bound family. It does not order the unpriced validation work.

\subsection{Complete inputs for the selected fractional clock}
The final clock uses the five lags $(.5,1,2,3,4)$ at width $.2$.
Table~\ref{tab:finalclockdirections} prints both exact rational directions as
terminating decimals. The first interval is $0\le\beta\le1$, and the second
is $1\le\beta\le\infty$. Their variance normalizations select lags $4$ and
$2$, respectively. Together with Table~\ref{tab:widthcounts} and the rejection
rule above, these specify the final 38,011,244-outcome clock design.
The extra constant coordinate in the shared comparison format has zero
coefficient in both directions.
\begin{table}[tbp]\centering\footnotesize
\begin{tabular}{crr}\toprule
Coordinate & $0\le\beta\le1$ & $1\le\beta\le\infty$\\\midrule
$x_g(.5)$ & 0.00045252797704605559 & 0\\
$x_g(1)$ & 0 & 0\\
$x_g(2)$ & -0.013403971453659395 & -0.21455490069471159\\
$x_g(3)$ & 0.0071429007619856462 & 0.084231573428110218\\
$x_g(4)$ & 0.10576250819996842 & 0\\
$y_g(.5)$ & 0 & -0.60854274385059426\\
$y_g(1)$ & 0.28603503828347343 & 0\\
$y_g(2)$ & 0.20885395577243296 & 0\\
$y_g(3)$ & 0.28619709426055256 & 0.086059052577326559\\
$y_g(4)$ & 0.092152003280881606 & 0.0066117294392574726\\
\bottomrule\end{tabular}
\caption{Exact directions of the selected clock, in fluctuation-then-response order. Both satisfy $\|z\|_1\le1$.}
\label{tab:finalclockdirections}\end{table}
The band supports used with these directions are the following exact
rationals. The tail allowance is added separately as in
equation~\eqref{eq:widthtail}.
\begin{gather*}
H_1=\frac{343901079704273405650528417112578754852082026440684591562660151203}{2854495385411919762116571938898990272765493248000000000000000000000}\\
H_2=\frac{-35086927402940869144836968488590953357166792044589494603516943367}{285449538541191976211657193889899027276549324800000000000000000000}.
\end{gather*}

\subsection{Visibility, control error and laboratory-time sensitivity}
For contrast $c$, phase half-angle $\phi$ and sensor coherence time $T_2$,
the reference floor follows from
\begin{equation}
 c>\frac{.9\exp[-T/8+(T+.18)h/T_2]}{\cos\phi},
 \qquad |\phi|<\pi/2.
 \label{eq:contrastthreshold}
\end{equation}
Here $T$ is the coherent-window duration in units of $h$: $T=\tau+d_j$ for two pulses and $T=d_j$ for a single pulse. With $\phi=.2$ and
$T_2=10^4h$, the shortest window $T=.2$ requires
$c>.895665965$. The reference condition alone would allow a much shorter
$T_2$; the finite-control comparison provides a separate requirement.
With a total deterministic unitary-channel error cap
$\delta_U=2\times10^{-5}$, the allocated complex-mean allowance is
\begin{equation}
 \sqrt2[.36h/T_2+\delta_U]<10^{-4},
 \qquad T_2/h>7099.097.
 \label{eq:controlthreshold}
\end{equation}
The $\delta_U$ cap is a channel-norm bound, rather than an average gate
infidelity. The reported example has $T_2/h\ge10^4$ and leaves the
remainder of the $10^{-3}$ residual cap for other unmatched errors.

These conditions describe how the example enters the stated power domain.
They do not measure optimal power outside that domain. A different
allocation of residual error or count design can be evaluated separately.
For $h=1\,\mu\mathrm s$, the acquisition-time upper bounds convert to approximately
39.88 minutes, 2.7575 hours and 10.7380 hours for the clock, full two-width
and added-wide-width-variance routes. Specification, clock and width validation
remain additional. A laboratory realization must meet all preparation, pulse,
coherence and control requirements at that same value of $h$.

\FloatBarrier
\begin{table}[!htbp]\centering\small
\begin{tabular}{rrrrr}\toprule
$c$ & $T_2/h$ & $\phi$ & Least reference margin & Control error bound\\\midrule
0.900 & 10000 & 0.20 & 0.00425 & 0.0000792\\
0.895 & 10000 & 0.20 & -0.00065 & 0.0000792\\
0.890 & 10000 & 0.20 & -0.00555 & 0.0000792\\
0.900 & 10000 & 0.25 & -0.00579 & 0.0000792\\
0.900 & 8000 & 0.20 & 0.00424 & 0.0000919\\
0.900 & 7000 & 0.20 & 0.00423 & 0.0001010\\
\bottomrule\end{tabular}
\caption{Sufficient conditions for the finite-control example, over the clock and width windows through $4.8h$. A negative reference margin or control error above $10^{-4}$ removes this implication for the chosen specifications. The statistical designs and their budgets are unchanged.}
\label{tab:modulesensitivity}\end{table}
\FloatBarrier

\subsection{Saved counts for the final separate-stream designs}
The final clock and full two-width allocations are tested with 16 independent
datasets each for the target, a common-$\beta=1.7$ two-mode state, and the
physical reflected thermal alias. This diagnostic uses exact finite-ramp
Gaussian coherences at $r=.01$, zero oscillator loss and zero sensor
dephasing, an allowed member of the certified class. The self phases are
computed by exponential integration of the trapezoid's three linear
segments. The two matched detector cohorts have gains $.91,.93$, offsets
$.005,-.004$ and phases $-.2,.2$. The full positive binary POVM retains
its transverse components and the separate conditional $I/Z$ coin laws.
The independent matrix/formula detector check differs by less than
$3.4\times10^{-16}$.

Each cohort is sampled once as a multinomial array over force label,
analysis phase, twirl coin and binary outcome, with master seed 20261008.
This is the aggregated law of the prescribed fresh choices and outcomes.
Every force-off reference and the 80,000-slot collective-variance stage is counted.
The arrays are saved before any decision and then supplied to a count-only
interface using 85-digit interval arithmetic. A separate standard-library
implementation imports neither the generator nor the decision code and agrees
with all 96 decisions; its largest score-margin difference is below
$2.5\times10^{-16}$.

The supplementary data contain the count arrays and fixed designs needed to
reproduce these decisions. Sections S6--S9 give the nonzero and predictably
changing loss guarantees and their final decision rule.
\FloatBarrier
\begin{table}[!htbp]\centering\small
\begin{tabular}{llrr}\toprule
Acquisition & State & Datasets & Rejections\\\midrule
Fractional clock & Target & 16 & 16\\
Fractional clock & Common $\beta=1.7$ & 16 & 0\\
Fractional clock & Reflected thermal alias & 16 & 0\\
Two widths & Target & 16 & 16\\
Two widths & Common $\beta=1.7$ & 16 & 0\\
Two widths & Reflected thermal alias & 16 & 0\\
\bottomrule\end{tabular}
\caption{Saved binary-count checks of the two final allocations. Each dataset uses its exact horizon, 38,011,244 or 158,404,641 outcomes. These finite diagnostic frequencies check the implementation; they do not establish the uniform 5\% error or 95\% power guarantees.}
\label{tab:finalcountchecks}\end{table}
\FloatBarrier
\section{Nonzero-loss overlap and fair comparisons}

\subsection{A constructive common observation law}

In the exact Gaussian map at unit strength write
$\ell_{E,s}=-Q_{E,s}+\ii(2S_E+sR_{E,\tau})$, where $E=T,A$ denotes target
or thermal alias. The commutator is stationary even for nonstationary
occupations, so every translated box has the same self phase $S_E$.
An unknown phase $c_E x$ per force box, with $x=\lambda^2$, is common to both
signs. Choose $c_T=0$, $c_A=S_T-S_A$. It matches the single-box self phase
and the two-box self phase at every strength. The remaining log difference is
\begin{equation}
 \Delta\ell_s=-(Q_{T,s}-Q_{A,s})+\ii s(R_{T,\tau}-R_{A,\tau}).
\end{equation}
For $m=\min(Q_{T,s},Q_{A,s})\ge0$, the exponential segment identity gives
\begin{equation}
 |A_T(x)-A_A(x)|\le\eta|\Delta\ell_s|x e^{-mx}.
\end{equation}
On $0\le x\le L$, the maximum of $xe^{-mx}$ is $Le^{-mL}$ if $mL\le1$
and $1/(em)$ otherwise. At $m=0$ it is $L$. Any certified lower bound
$0\le m_L\le m$ can replace $m$, because the maximum decreases with $m$.
This proves a continuum-strength bound without a strength grid. The first-box
single-box collective-variance mean difference is exactly
$\eta_0|e^{-V_T/2}-e^{-V_A/2}|$ after the same phase shift. References agree.

The comparison uses $d=0.2$, integer lags $1,\ldots,5$,
$L=1.1$, loss-bath occupation zero and the physical three-mode alias from S2.
At $\kappa h=8\times10^{-5}$, the maximal signal upper bound is
$2.40109\times10^{-5}$ for gain $1$, and $2.16098\times10^{-5}$ for gain
$0.9$. The respective single-box bounds are below $5.4440\times10^{-7}$ and
$4.8996\times10^{-7}$. These satisfy either $10^{-4}$ or $10^{-3}$ residual
caps. At $\kappa h=4\times10^{-4}$ the analogous signal bounds exceed
$10^{-4}$ but satisfy $10^{-3}$; failure of the smaller sufficient bound is
inconclusive for intersection. Across six rate/gain pairs and two caps, ten
cases pass and two are inconclusive. Independent 100-digit reconstruction
checks 453 stored intervals and includes four direct kernel-integral anchors.
It explicitly confirms that the nonzero-loss ideal generators differ.

For the $S$ sequence, the same signal applies. The $D$ sequence is the signed-minus
signal after multiplication by $e^{-2\ii\lambda^2S_T}$, common to both models
after the nuisance-phase choice. Modulus differences are therefore unchanged.
Let $M(L)=\max_{0\le x\le L}xe^{-mx}$. For $L_2\ge L_1$, writing
$x=(L_2/L_1)y$ gives $xe^{-mx}\le(L_2/L_1)ye^{-my}$ for $0\le y\le L_1$.
Thus $M(1.7)\le(17/11)M(1.1)$. Multiplying the saved signal bounds by $17/11$
covers every strength in the implemented unsplit pulse family without another grid. At
$\kappa h=8\times10^{-5}$ and gain $0.9$ the bound is
$3.339691292123\times10^{-5}<10^{-3}$; the single-box bound is unchanged.

Now adopt the \emph{entire} fixed setting-dependent complex-mean error envelope
as the allowed observation class. For every permitted setting $k$ choose the
null residual $\delta_A(k)=A_T(k)-A_A(k)$ and target residual zero. At a Pauli
angle $\phi$, the null and target binary means both equal
$\Re[e^{-\ii\phi}A_T(k)]$. The probability is $(1+\mu)/2$ and is physical
because it equals the target's. At opposite analysis phases the residual
changes sign, matching the complete labeled binary record, not merely a
compressed statistic. For the detector-twirl implementation choose the same
allowed diagonal detector for both models. Each coin value then has the same
conditional output kernel, so even retained coin records have a common law.
Equality of averaged laws alone would not justify this claim for an arbitrary
fixed anisotropic detector with accessible coin records. The same construction covers collective-variance and reference data.
It is a fixed function of controls before data collection.

Fresh independent preparations then give the same conditional binary kernel
at every history. An adaptive algorithm choosing among these controls has
identical joint data and action laws under the pair, by induction; a stopping
or decision functional has the same law as well. A level-$\alpha$ test over
this observation class must therefore have power at most $\alpha$ on the
paired unbiased alternative, at any shot count. A bound on a narrower physical
electronics model is not enough to infer that our constructed residual belongs
to it. This additional minimax premise is essential. A calibrated self phase,
other pulse widths, exact bare variance, longer integer lags under known loss,
mode-resolved measurements or unmodeled detector records are outside the theorem.

\begin{figure}[tbp]
\centering\includegraphics[width=\linewidth]{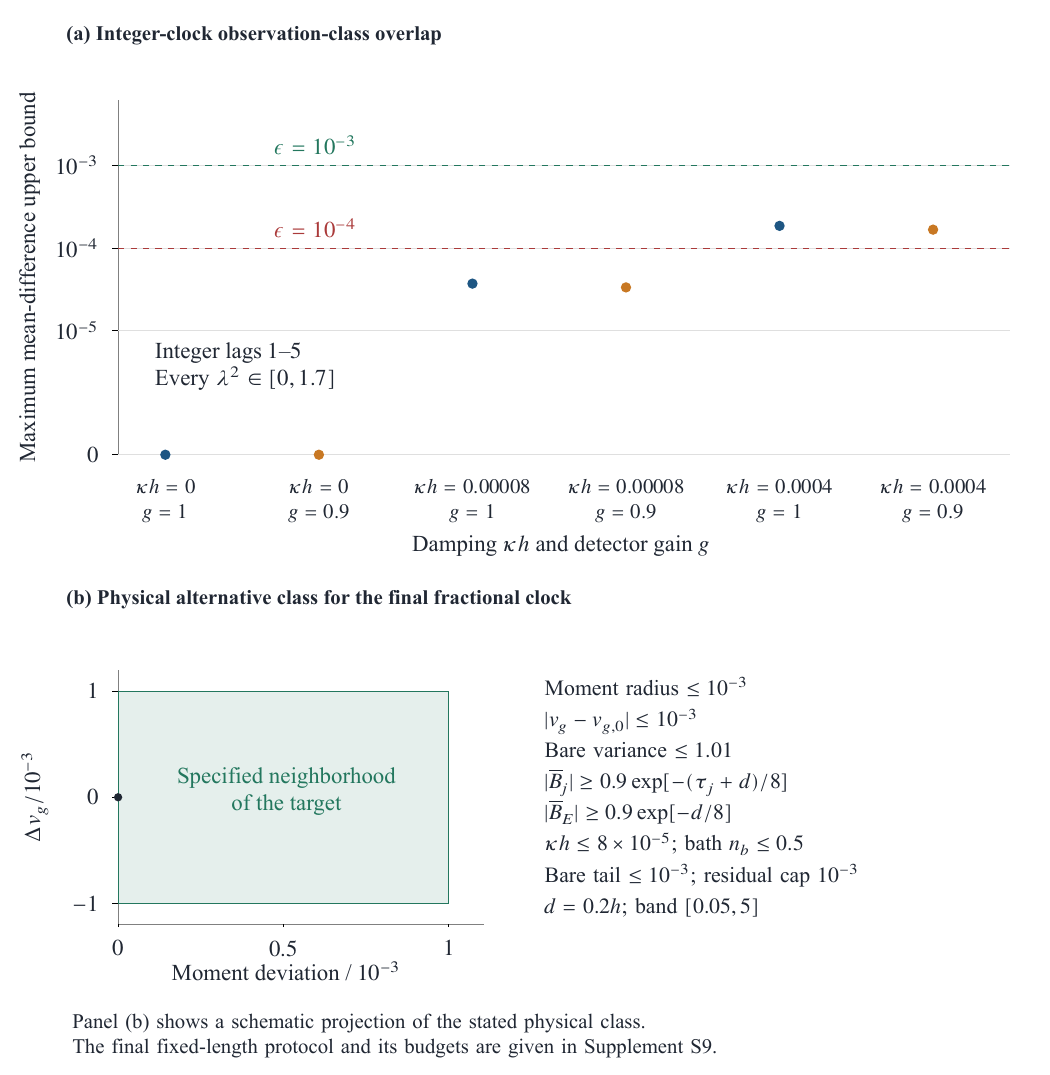}
\caption{Two uses of an error allowance. (a) The frozen Gaussian-loss certificates bound the target--alias complex-mean difference at integer lags $1,\ldots,5$ and throughout $0\le\lambda^2\le1.7$, using the proved $17/11$ extension of the original strength bound. A point below a residual cap constructs a common allowed observation law; a point above it leaves intersection unresolved. Zero loss has an exact physical alias. At nonzero loss the construction uses the full setting-dependent mean-error class and the same diagonal detector, including its randomization coins. (b) The rectangle projects the stated physical alternative class onto moment deviation and filtered variance. Its reference, loss, tail and residual-error requirements are listed alongside. The final fractional-clock protocol in S9 fixes the allocation and gives the 38,011,244-outcome guarantee. The rectangle's area carries no statistical probability, and not every plotted coordinate pair need be physically realized.}
\end{figure}

\subsection{An isolated-mode sideband baseline with a measured detector reference}

For one thermal oscillator $p_n=(1-r)r^n$, $r=e^{-\beta\omega}$, let $f_n$
be the matched reciprocal Fock-pair transition probability. Then
\begin{equation}
 P_b=\sum_{n\ge0}p_nf_n,\qquad P_r=\sum_{n\ge0}p_{n+1}f_n=rP_b.
\end{equation}
This identity does not require weak pulse area. We use the isolated single-spin
Jaynes--Cummings/Lamb--Dicke expression $f_n=\sin^2[\theta\sqrt{n+1}/2]$.
Mode isolation and matched red/blue pulses remain assumptions; global many-spin
excitation requires its own model\cite{Turchette2000,Vybornyi2023}.

Allow a fixed common binary detector with unknown offset $a$ and positive gain
$g$: $b_i=a+gP_{b,i}$, $d_i=a+gP_{r,i}$. Measure an independent ground-probe
reference of probability $a$. Five settings each use $N_{\rm sb}$ trials.
For per-probability systematic cap $\epsilon_p$, simultaneous coverage uses
\begin{equation}
 h_N=\sqrt{\log(10/\alpha)/(2N_{\rm sb})}+\epsilon_p.
\end{equation}
With $\widehat a$ measured, require positive lower denominators and form
\begin{equation}
 r_{L,i}=\max\!\left(0,\frac{\widehat d_i-\widehat a-2h_N}{\widehat b_i-\widehat a+2h_N}\right),\quad
 r_{U,i}=\min\!\left(1,\frac{\widehat d_i-\widehat a+2h_N}{\widehat b_i-\widehat a-2h_N}\right).
\end{equation}
Empty intersections, invalid denominators or overlapping temperature intervals
are inconclusive. For supplied $0<\omega_{L,i}\le\omega_i\le\omega_{U,i}$,
\begin{equation}
 \beta_{L,i}=-\log r_{U,i}/\omega_{U,i},\qquad
 \beta_{U,i}=-\log r_{L,i}/\omega_{L,i}.
\end{equation}
The upper endpoint is infinite when $r_L=0$. Disjoint intervals reject a common
temperature with type-I error at most $\alpha=.05$, conditional on the physical
frequency/error premises. The count API never receives the true $a,g,\beta$ or
transition probability.

For target power at $g\ge g_{\min}=.9$ and arbitrary physical offset, worst
ratios in the needed direction are
\begin{equation}
 r_{1,U}^*=\frac{g_{\min}P_{r,1}+4h_N}{g_{\min}P_{b,1}-4h_N},\qquad
 r_{2,L}^*=\frac{g_{\min}P_{r,2}-4h_N}{g_{\min}P_{b,2}+4h_N}.
\end{equation}
The factor four accounts for both signal/reference uncertainty and displacement of
observed centers. Their dependence on $g$ makes $g_{\min}$ conservative.
Require positive denominators, ratios in $(0,1)$ and
$-\log r_{1,U}^*/\omega_{U,1}>-\log r_{2,L}^*/\omega_{L,2}$.
Only this design calculation uses target probabilities. One calculation uses
400 positive blue terms with tail $[0,r^{400}]$; an independent 600-term
calculation uses $(1-\cos)/2$ and a direct red sum. Thus no finite Fock cutoff is exact,
and the arbitrary-temperature null theorem uses the exact ratio identity.

Four areas per mode, $\theta/\pi\in\{.5,1,1.5,2\}$, give 16 pairs per bias cap.
The selected pair is $(\pi,\pi/2)$ in both cap families. With $\pm.1\%$ frequency
intervals, the sufficient counts after a $1.1$ safety multiplier are
$N_{\rm sb}=2243$ for $\epsilon_p=10^{-4}$ and $2275$ for
$\epsilon_p=10^{-3}/(2\sqrt2)$. They cost $5N_{\rm sb}=11215$ and $11375$
probe preparations, of which $4N_{\rm sb}$ require fresh baths and $N_{\rm sb}$
are detector references. The second cap has a worst-case balanced complex-mean
interpretation of $10^{-3}$; it is not an identical microscopic error model.
Eight rounded-exact-mean diagnostics reject the targets and leave common
temperatures and very small gain inconclusive. They are not binomial samples.

Complete sideband time is $N_{\rm sb}(c_{b1}+c_{r1}+c_{b2}+c_{r2}+c_{\rm ref})$
plus frequency, addressing, pulse-matching and bias validation/refresh costs.
The pulse areas do not determine a physical duration without an allowed Rabi
rate and isolation criterion. This additional access must be admitted whenever
available; its favorable result bounds the use case of collective certification.

\section{Negative results and limits of task transfer}

\subsection{One unknown-temperature lag cannot reject the finite mixture}

On a nonfolding band $0<\omega h<\pi$, a finite positive-temperature mixture
has $0<y<\sqrt{1-x^2}$. The strict upper inequality follows from
$\gam_i<1$ and concavity of $\sqrt{1-x^2}$ on the cosine interval. The frequency
$\omega_*=(\arccos x)/h$ lies in the original band and
\begin{equation}
 \beta_*^{(1)}=\frac{2}{\omega_*}
                   \atanh\frac{y}{\sqrt{1-x^2}}
\end{equation}
is finite and positive. One mode realizes both observations. For the main
point two-mode example, $\omega_*\simeq1.19413$ and $\beta_*^{(1)}\simeq0.539398$.
The two-lag certificate therefore establishes minimality by the number of
complex point lags for this example, not by shots, arbitrary filters or settings.

\subsection{Temperature information can leave a later task unchanged}

For any fixed probability measure $q$ on the strict nonfolding band with the
observed cosine mean, define
$G_q(\beta)=\int\tanh(\beta\omega/2)\sin(\omega h)\dd q$.
It increases strictly from zero to $\int\sin(\omega h)\dd q$. The concave
curve $\sqrt{1-x^2}$ lies above its band-endpoint chord $L_\infty(x)$, so the
last integral is at least that chord at the observed $x$. If
$0<y<L_\infty(x)$, every cosine-compatible $q$ has a unique finite $\beta$
solving $G_q(\beta)=y$. Hence the set of allowed variance spectra is unchanged
when the response is added and temperature remains free.

For fixed initial variance, any later functional depending only on $q$ has
the same compatible range. For example, a constant pure-dephasing coupling
$H_{\rm int}=(\sigma_z/2)B$ gives
\begin{equation}
 |L(t)|=\exp\left[-v\int\frac{1-\cos\omega t}{\omega^2}\dd q(\omega)\right].
\end{equation}
This negative result is different from a fixed-known-temperature positive
example. The strict inequality excludes boundary cases that can remove
particular spectra; failure of the sufficient inequality alone does not prove
a useful prediction improvement.
If only one later output is needed, measuring that output
directly remains a required comparator.


\begin{thebibliography}{99}
\raggedright
\setlength{\itemsep}{2pt}
\bibitem{PazSilva2017} G. A. Paz-Silva, L. M. Norris and L. Viola, Multiqubit spectroscopy of Gaussian quantum noise, \textit{Phys. Rev. A} \textbf{95}, 022121 (2017). \href{https://doi.org/10.1103/PhysRevA.95.022121}{doi:10.1103/PhysRevA.95.022121}.

\bibitem{Wang2021} Y.-X. Wang and A. A. Clerk, Intrinsic and induced quantum quenches for enhancing qubit-based quantum noise spectroscopy, \textit{Nat. Commun.} \textbf{12}, 6528 (2021). \href{https://doi.org/10.1038/s41467-021-26868-7}{doi:10.1038/s41467-021-26868-7}.

\bibitem{Roman2020} R. Rom\'an-Ancheyta, B. \c{C}akmak and \"O. E. M\"ustecapl\i o\u{g}lu, Spectral signatures of non-thermal baths in quantum thermalization, \textit{Quantum Sci. Technol.} \textbf{5}, 015003 (2020). \href{https://doi.org/10.1088/2058-9565/ab5e4f}{doi:10.1088/2058-9565/ab5e4f}.

\bibitem{Karlsson2013} J. Karlsson and T. T. Georgiou, Uncertainty bounds for spectral estimation, \textit{IEEE Trans. Autom. Control} \textbf{58}, 1659--1673 (2013). \href{https://doi.org/10.1109/TAC.2013.2251775}{doi:10.1109/TAC.2013.2251775}.

\bibitem{Yang2018} Z. Yang and L. Xie, Frequency-selective Vandermonde decomposition of Toeplitz matrices with applications, \textit{Signal Process.} \textbf{142}, 157--167 (2018). \href{https://doi.org/10.1016/j.sigpro.2017.07.024}{doi:10.1016/j.sigpro.2017.07.024}.

\bibitem{Frame2021} T. Chalermpusitarak, B. Tonekaboni, Y. Wang, L. M. Norris, L. Viola and G. A. Paz-Silva, Frame-based filter-function formalism for quantum characterization and control, \textit{PRX Quantum} \textbf{2}, 030315 (2021). \href{https://doi.org/10.1103/PRXQuantum.2.030315}{doi:10.1103/PRXQuantum.2.030315}.

\bibitem{Ferrie2018} C. Ferrie, C. Granade, G. A. Paz-Silva and H. M. Wiseman, Bayesian quantum noise spectroscopy, \textit{New J. Phys.} \textbf{20}, 123005 (2018). \href{https://doi.org/10.1088/1367-2630/aaf207}{doi:10.1088/1367-2630/aaf207}.

\bibitem{Shitara2025} N. Shitara and A. Montoya-Castillo, Fast, accurate, and error-resilient variational quantum noise spectroscopy, \textit{J. Chem. Phys.} \textbf{164}, 061101 (2026). \href{https://doi.org/10.1063/5.0312403}{doi:10.1063/5.0312403}.

\bibitem{Fawzi2024} H. Fawzi, O. Fawzi and S. O. Scalet, Certified algorithms for equilibrium states of local quantum Hamiltonians, \textit{Nat. Commun.} \textbf{15}, 7394 (2024). \href{https://doi.org/10.1038/s41467-024-51592-3}{doi:10.1038/s41467-024-51592-3}.

\bibitem{Artymowicz2024} A. Artymowicz, H. Fawzi, O. Fawzi and S. O. Scalet, Certified algorithms for quantum Hamiltonian learning via energy-entropy inequalities, \href{https://arxiv.org/abs/2410.23284}{arXiv:2410.23284v1} (2024).

\bibitem{Maciejewski2009} M. W. Maciejewski, H. Z. Qui, I. Rujan, M. Mobli and J. C. Hoch, Nonuniform sampling and spectral aliasing, \textit{J. Magn. Reson.} \textbf{199}, 88--93 (2009). \href{https://doi.org/10.1016/j.jmr.2009.04.006}{doi:10.1016/j.jmr.2009.04.006}.

\bibitem{Norris2018} L. M. Norris, D. Lucarelli, V. M. Frey, S. Mavadia, M. J. Biercuk and L. Viola, Optimally band-limited spectroscopy of control noise using a qubit sensor, \textit{Phys. Rev. A} \textbf{98}, 032315 (2018). \href{https://doi.org/10.1103/PhysRevA.98.032315}{doi:10.1103/PhysRevA.98.032315}.

\bibitem{Digital2024} G. Wang, Y. Zhu, B. Li, C. Li, L. Viola, A. Cooper and P. Cappellaro, Digital noise spectroscopy with a quantum sensor, \textit{Quantum Sci. Technol.} \textbf{9}, 035006 (2024). \href{https://doi.org/10.1088/2058-9565/ad3846}{doi:10.1088/2058-9565/ad3846}.

\bibitem{Klein2007} A. Klein, M. Bruderer, S. R. Clark and D. Jaksch, Dynamics, dephasing and clustering of impurity atoms in Bose--Einstein condensates, \textit{New J. Phys.} \textbf{9}, 411 (2007). \href{https://doi.org/10.1088/1367-2630/9/11/411}{doi:10.1088/1367-2630/9/11/411}.

\bibitem{Berg2022} E. van den Berg, Z. K. Minev and K. Temme, Model-free readout-error mitigation for quantum expectation values, \textit{Phys. Rev. A} \textbf{105}, 032620 (2022). \href{https://doi.org/10.1103/PhysRevA.105.032620}{doi:10.1103/PhysRevA.105.032620}.

\bibitem{Proctor2020} T. Proctor, M. Revelle, E. Nielsen, K. Rudinger,
D. Lobser, P. Maunz, R. Blume-Kohout and K. Young,
Detecting and tracking drift in quantum information processors,
\textit{Nat. Commun.} \textbf{11}, 5396 (2020).
\href{https://doi.org/10.1038/s41467-020-19074-4}{doi:10.1038/s41467-020-19074-4}.

\bibitem{Khan2024} M. Q. Khan, W. Dong, L. M. Norris and L. Viola,
Multiaxis quantum noise spectroscopy robust to errors in state preparation
and measurement, \textit{Phys. Rev. Applied} \textbf{22}, 024074 (2024).
\href{https://doi.org/10.1103/PhysRevApplied.22.024074}{doi:10.1103/PhysRevApplied.22.024074}.

\bibitem{Freedman1975} D. A. Freedman, On tail probabilities for martingales,
\textit{Ann. Probab.} \textbf{3}, 100--118 (1975).
\href{https://doi.org/10.1214/aop/1176996452}{doi:10.1214/aop/1176996452}.

\bibitem{Jin2026} Y.-D. Jin, Z.-F. Ye and W.-L. Ma, Resource-Efficient Noise Spectroscopy for Generic Quantum Dephasing Environments, \href{https://arxiv.org/abs/2601.18290v1}{arXiv:2601.18290v1} (2026).

\bibitem{Turchette2000} Q. A. Turchette \textit{et al.}, Heating of trapped ions from the quantum ground state, \textit{Phys. Rev. A} \textbf{61}, 063418 (2000). \href{https://doi.org/10.1103/PhysRevA.61.063418}{doi:10.1103/PhysRevA.61.063418}.

\bibitem{Vybornyi2023} I. Vybornyi, L. S. Dreissen, D. Kiesenhofer, H. Hainzer, M. Bock, T. Ollikainen, D. Vadlejch, C. F. Roos, T. E. Mehlst\"aubler and K. Hammerer, Sideband thermometry of ion crystals, \textit{PRX Quantum} \textbf{4}, 040346 (2023). \href{https://doi.org/10.1103/PRXQuantum.4.040346}{doi:10.1103/PRXQuantum.4.040346}.

\bibitem{Rasmusson2021} A. J. Rasmusson, M. D'Onofrio, Y. Xie, J. Cui and P. Richerme, Optimized pulsed sideband cooling and enhanced thermometry of trapped ions, \textit{Phys. Rev. A} \textbf{104}, 043108 (2021). \href{https://doi.org/10.1103/PhysRevA.104.043108}{doi:10.1103/PhysRevA.104.043108}.

\end{thebibliography}
\end{document}